\documentstyle[12pt]{article}

\def\thebibliography#1{\section*{}\list
{$^{\arabic{enumi}}$}{\settowidth\labelwidth{#1}\leftmargin\labelwidth
\advance\leftmargin\labelsep
\usecounter{enumi}}
\def\newblock{\hskip .11em plus .33em minus .07em}
\sloppy\clubpenalty4000\widowpenalty4000
\sfcode`\.=1000\relax}

\def\op#1{\mathop{\fam0 #1}\limits}

\newcommand{\id}{{\rm Id\,}}
\newcommand{\pr}{{\rm pr}}
\newcommand{\Ker}{{\rm Ker\,}}
\newcommand{\nm}[1]{\mid {#1}\mid}

\newcommand{\beq}{\begin{equation}}
\newcommand{\eeq}{\end{equation}}
\newcommand{\ben}{\begin{eqnarray}}
\newcommand{\een}{\end{eqnarray}}
\newcommand{\be}{\begin{eqnarray*}}
\newcommand{\ee}{\end{eqnarray*}}
\newcommand{\bea}{\begin{eqalph}}
\newcommand{\eea}{\end{eqalph}}

\newcommand{\cA}{{\cal A}}
\newcommand{\cT}{{\cal T}}
\newcommand{\cP}{{\cal P}}
\newcommand{\cR}{{\cal R}}
\newcommand{\cL}{{\cal L}}
\newcommand{\cV}{{\cal V}}
\newcommand{\cE}{{\cal E}}
\newcommand{\cH}{{\cal H}}
\newcommand{\cF}{{\cal F}}
\newcommand{\cS}{{\cal S}}

\newcommand{\cJ}{{\cal J}}
\newcommand{\bL}{{\bf L}}
\newcommand{\bR}{{\bf R}}
\newcommand{\bC}{{\bf C}}
\newcommand{\bZ}{{\bf Z}}
\newcommand{\al}{\alpha}
\newcommand{\bt}{\beta}

\newcommand{\la}{\lambda}
\newcommand{\La}{\Lambda}
\newcommand{\f}{\phi}
\newcommand{\om}{\omega}
\newcommand{\Om}{\Omega}
\newcommand{\m}{\mu}

\newcommand{\g}{\gamma}
\newcommand{\G}{\Gamma}

\newcommand{\ve}{\varepsilon}
\newcommand{\th}{\theta}

\newcommand{\vt}{\vartheta}
\newcommand{\si}{\sigma}
\newcommand{\w}{\wedge}
\newcommand{\wt}{\widetilde}
\newcommand{\wh}{\widehat}
\newcommand{\ol}{\overline}
\newcommand{\dr}{\partial}
\newcommand{\ar}{\op\longrightarrow}
\newcommand{\llra}{\longleftrightarrow}
\newcommand{\ot}{\otimes}
\newcommand{\ap}{\approx}
\newcommand{\nw}[1]{[{#1}]}
\newcommand{\der}{\rm Der}

\let\ssection=\section
\renewcommand{\section}{\setcounter{equation}{0}\ssection}

\newcounter{eqalph}
\newcounter{equationa}
\newcounter{theorem}
\newcounter{proposition}
\newcounter{lemma}
\newcounter{corollary}
\newcounter{definition}
\newcounter{remark}

\newenvironment{eqalph}{\stepcounter{equation}
\setcounter{equationa}{\value{equation}}
\setcounter{equation}{0}

\begin{eqnarray}}{\end{eqnarray}\setcounter{equation}{\value{equationa}}}

\def\theremark{\arabic{remark}}

\newenvironment{proof}{\noindent 
{\it Proof:}}{\medskip}
\newenvironment{rem}{\refstepcounter{remark}\medskip\noindent{\it
Remark \theremark:}}{\medskip}
\newenvironment{theo}{\refstepcounter{definition} 
\medskip\noindent{\it Theorem \thedefinition:}}{\medskip}
\newenvironment{prop}{\refstepcounter{definition} 
\medskip\noindent{\it Proposition \thedefinition:}}{\medskip}
\newenvironment{lem}{\refstepcounter{definition} 
\medskip\noindent{\it Lemma \thedefinition:}}{\medskip}
\newenvironment{cor}{\refstepcounter{definition} 
\medskip\noindent{\it Corollary \thedefinition:}}{\medskip}

\begin{document}
\hbox{}

{\parindent=0pt

{\large\bf Covariant Hamiltonian field theory}
\bigskip

{\sc G.Giachetta and L.Mangiarotti}\footnote{Electronic mail:
mangiaro@camserv.unicam.it}

{\sl Department of Mathematics and Physics, University of Camerino, 62032
Camerino (MC), Italy}
\medskip

{\sc G. Sardanashvily}\footnote{Electronic mail:
sard@grav.phys.msu.su}

{\sl Department of Theoretical Physics, 
Moscow State University, 117234 Moscow, Russia}
\bigskip
 
We study the relations between the equations of first order Lagrangian field
theory on fiber bundles and the covariant Hamilton equations on the
finite-dimensional polysymplectic phase space of covariant Hamiltonian
field theory. The main peculiarity of these Hamilton equations lies in the fact
that, for degenerate systems, they contain additional gauge fixing conditions.
We develop the BRST extension of the covariant
Hamiltonian formalism, characterized by a Lie superalgebra of
BRST and anti-BRST symmetries.
\bigskip

 }
\bigskip 

\noindent 
{\bf I. INTRODUCTION}
\bigskip

 As is well known, when applied to field theory, the familiar
symplectic techniques of mechanics take the form of instantaneous Hamiltonian
formalism on an infinite-dimensional phase space. 
The finite-dimensional covariant Hamiltonian approach to field theory is
vigorously developed from the seventies in its multisymplectic and
polysymplectic variants.$^{1-3}$ Its final purpose 
is the covariant Hamiltonian quantization of field theory.

In the framework of this approach, one deals with the
following types of PDEs:  Euler--Lagrange and
 Cartan equations in the Lagrangian formalism, Hamilton--De Donder
equations in multisymplectic Hamiltonian formalism, covariant Hamilton
equations and  restricted Hamilton equations in polysymplectic
Hamiltonian formalism. If a Lagrangian is
hyperregular, all these PDEs are equivalent. 
The present work
addresses  degenerate semiregular and almost
regular Lagrangians. From the mathematical viewpoint, these notions of
degeneracy are particularly appropriate in order to study the relations between
the above-mentioned PDEs. From the physical one, Lagrangians of almost all
field theories are of these types. 

To formulate our results, let us recall briefly some notions.
Given a fiber bundle
$Y\to X$, coordinated by $(x^\la,y^i)$, a first order Lagrangian $L$ is
defined as a horizontal  density
\beq
L=\cL\om: J^1Y\to\op\w^nT^*X, \quad \om=dx^1\w\cdots dx^n, \quad n=\dim X,
\label{cmp1}
\eeq
on the affine jet bundle $J^1Y\to Y$, provided with the adapted coordinates
$(x^\la,y^i,y^i_\la)$. $J^1Y$ can be seen as a  finite-dimensional
configuration space of fields represented by sections of $Y\to X$. 

Given a Lagrangian
$L$ (\ref{cmp1}), the associated Euler--Lagrange equations 
define both the equations of the variational
problem on $Y$ for $L$ and the kernel of the
Euler--Lagrange operator, which can be also introduced in an intrinsic
way as a coboundary element of the variational cochain complex. 

The Cartan equations characterize 
the variational problem on $J^1Y$ for the
Poincar\'e--Cartan form $H_L$, which is a horizontal Lepagean equivalent of
$L$ on $J^1Y\to Y$, i.e. $L=h_0(H_L)$, where $h_0$ is the horizontal
projection (\ref{cmp100}). At the same time, the Cartan equations can be seen
both as the kernel of the Euler--Lagrange--Cartan operator and the Hamilton
equations of the Lagrangian polysymplectic structure on
$J^1Y$. The Cartan equations are the Lagrangian
counterpart of covariant Hamilton equations.

The Hamilton-De Donder equations  and the covariant Hamilton
equations are related to two different Legendre morphisms in the first
order calculus of variations. 

Firstly, every
Poincar\'e--Cartan form
$H_L$ yields the Legendre morphism
$\wh H_L$ of
$J^1Y$ to the homogeneous Legendre bundle
\beq
Z_Y=J^{1\star}Y = T^*Y\w(\op\w^{n-1}T^*X)  \label{N41}
\eeq
which is the affine $\op\w^{n+1}$-valued dual of $J^1Y\to Y$,$^{1,4}$
and is treated as
a homogeneous finite-dimensional phase space of fields.
$Z_Y$ is provided with the canonical
exterior $n$-form
$\Xi_Y$ (\ref{N43}) and the multisymplectic form $d\Xi_Y$. If 
$\wh H_L(J^1Y)$ is an imbedded subbundle of $Z_Y\to Y$, the
pull-back of $\Xi_Y$ yields the Hamilton--De Donder
equations on $\wh H_L(J^1Y)$. 
If a Lagrangian 
$L$ is almost regular, these equations are quasi-equivalent to the
Cartan equations, i.e., there is a surjection of the set of solutions
of the Cartan equations onto that of the Hamilton-De Donder equations.$^1$ 

Secondly, every Lagrangian $L$ defines the Legendre map $\wh L$ of
$J^1Y$ to the Legendre bundle 
\beq
\Pi=\op\w^nT^*X\op\ot_YV^*Y\op\ot_YTX, \label{00}
\eeq
provided with the holonomic coordinates $(x^\la,y^i,p^\la_i)$, which can be
seen as a finite-dimensional momentum phase space of fields.$^{5-9}$
The relationship  between multisymplectic and polysymplectic phase spaces is
given by the exact sequence
\beq
0\ar\Pi\op\times_X\op\w^nT^*X\hookrightarrow Z_Y\ar\Pi\ar 0, \label{b417}
\eeq
where 
\beq
\pi_{Z\Pi}:Z_Y\to \Pi\label{b418'}
\eeq
is a 1-dimensional affine bundle. Given any section
$h$ of $Z_Y\to\Pi$, the pull-back 
\beq
 H=h^*\Xi_Y= p^\la_i dy^i\w \om_\la -\cH\om \label{b418}
\eeq
is a
polysymplectic Hamiltonian form on $\Pi$.$^{2-4}$
The Legendre $\Pi$ is equipped with the
canonical polysymplectic form $\Om_Y$ (\ref{406}).$^{2,3}$ 
This form differs from those in
Refs. [4-6], and is globally defined. With the polysymplectic
form
$\Om_Y$, one introduces Hamiltonian connections and covariant Hamilton
equations 
\bea
&& y^i_\la=\dr^i_\la\cH, \label{b4100a}\\
&& p^\la_{\la i}=-\dr_i\cH,\label{b4100b}
\eea
which are PDEs on the phase space $\Pi$ defined by the kernel of the Hamilton
operator $\cE_H$ (\ref{3.9}). If
$X=\bR$, covariant Hamiltonian formalism provides the adequate
Hamiltonian formulation of time-dependent mechanics.$^{10,11}$
In this case, $Z_Y=T^*Y$ and $\Pi=V^*Y$ are the
homogeneous and momentum phase spaces
of time-dependent mechanics, respectively.

It should be emphasized that
a Hamiltonian form $H$ (\ref{b418}) is the Poincar\'e--Cartan form of the
Lagrangian
\beq
L_H=h_0(H) = (p^\la_iy^i_\la - \cH)\om \label{Q3}
\eeq
on the jet manifold $J^1\Pi$. It is the 
the Poincar\'e--Cartan form (\ref{303})
of this Lagrangian. It follows that the Euler--Lagrange operator (\ref{305})
for $L_H$ is precisely the Hamilton operator $\cE_H$ (\ref{3.9}) for
$H$ and, consequently, the Euler--Lagrange equations for
$L_H$ are equivalent to the Hamilton equations for $H$. 
The Lagrangian $L_H$ plays a prominent role
in the path integral approach to quantization of Hamiltonian
systems.$^{12-14}$

The results of this paper demonstrate that polysymplectic Hamiltonian formalism
is not equivalent to the Lagrangian one, but can provide the adequate
description of degenerate field  systems which do not necessarily possess 
gauge symmetries. 

We show that, if $r:X\to\Pi$ is a solution of the Hamilton equations for a
Hamiltonian form $H$ associated with a semiregular Lagrangian $L$ and if $r$
lives in the Lagrangian constraint space $\wh L(J^1Y)$, then the
projection of $r$ onto $Y$ is a solution of the Euler--Lagrange
equations for $L$. The converse assertion is more intricate.
One needs a complete set of associated Hamiltonian forms
in order to exhaust all solutions of
the Euler--Lagrange equations for a degenerate Lagrangian. 
It follows that the
covariant Hamilton equations contain additional conditions
in comparison with the Euler--Lagrange ones. In the case of 
almost regular Lagrangians, one can introduce the constrained
Hamilton equations.  They are weaker than the Hamilton equations
restricted to  the Lagrangian constraint space, are 
equivalent to the Hamilton--De Donder equations and, consequently, are
quasi-equivalent to the Cartan equations.

The detailed analysis  
of degenerate quadratic Lagrangian systems in Section VI is appropriate
for application to many physical models. We
find a complete set of associated Hamiltonian forms. The
key point  is the splitting of the configuration space $J^1Y$ into the dynamic
sector and the gauge one coinciding with the kernel of the Legendre map
$\wh L$. As an immediate consequence of this splitting, one can separate a part
of the Hamilton equations independent of
momenta which play the role of gauge-type conditions, while other
equations restricted to the Lagrangian constraint space coincide with the
constrained Hamilton equations, and are quasi-equivalent to the Cartan
equations. 

Thus, we observe that the main features in gauge
theory are not directly related to the gauge invariance condition, but are
common in all field models with degenerate quadratic Lagrangians. The
important peculiarity of the Hamiltonian description of these models lies in
the fact that, in comparison with a Lagrangian, an associated Hamiltonian
form $H$ and the Lagrangian $L_H$ (\ref{Q3}) contain gauge fixing terms.
Therefore we will construct the BRST extension of the Hamiltonian form $H$
(\ref{b418}) and the Lagrangian $L_H$ (\ref{Q3}) in order to provide them with
symmetries which lead, e.g., to the corresponding Slavnov identities under
quantization. This is a preliminary step towards the covariant Hamiltonian
quantization of degenerate systems. 

A natural idea of the covariant Hamilton quantizations is also to generalize
the Poisson bracket in symplectic mechanics to multisymplectic or
polysymplectic manifolds and then to quantize it.$^{15}$ The main difficulty is
that the bracket must be globally defined. Let us note that
multisymplectic manifolds such as $(Z_Y,d\Xi_Y)$,
look rather promising for algebraic constructions since multisymplectic forms
are exterior forms.$^{16-18}$  

Nevertheless, the above mentioned $X=\bR$ reduction of the covariant
Hamiltonian formalism leads to time-dependent mechanics, but not conservative
symplectic mechanics.
In this case, the momentum phase space
$V^*Y$, coordinated by $(t,y^i,p_i=\dot y_i)$, is endowed with the canonical
degenerate Poisson structure given by the bracket
\beq
\{f,g\}_V=\dr^if\dr_i g-\dr^ig\dr_i f, \qquad f,g\in C^\infty(V^*Y).
\label{c3}
\eeq
However, the Poisson bracket $\{\cH,f\}_V$ of a Hamiltonian $\cH$ and
functions $f$ on the momentum phase space $V^*Y$ fails to be a well-behaved
entity because $\cH$ is not a scalar
with respect to time-dependent transformations. In particular, the equality
$\{\cH,f\}_V=0$ is not preserved under such
transformations.$^{10,11}$ As a consequence, the evolution equation in
time-dependent mechanics is not reduced to a Poisson bracket. At the same
time, the Poisson bracket (\ref{c3}) leads to the following current algebra
bracket. Let
$u=u^i\dr_i$ be a vertical vector field on $Y\to\bR$, and $J_u=u^ip_i$ the
corresponding symmetry current on $V^*Y$ along $u$. The symmetry
currents
$J_u$ constitute a Lie algebra with respect to the bracket
\be
[J_u,J_{u'}]=\{J_u,J_{u'}\}_V= J_{[u,u']}.
\ee
This current algebra bracket can be extended to the general polysymplectic case
as follows.

There is the canonical isomorphism
\be
\ol\th= p^\la_i\ol dy^i\w\om_\la: \Pi\to
V^*Y\op\w_Y(\op\w^{n-1}T^*X).
\ee
Let $u=u^i\dr_i$ be a vertical vector field on
$Y\to X$. The corresponding symmetry current (\ref{cmp129})
is a horizontal exterior $(n-1)$-form
\beq
J_u=u\rfloor\ol\th=u^ip_i^\la\om_\la \label{c1}
\eeq
on the Legendre bundle $\Pi$ (\ref{00}). The symmetry currents (\ref{c1})
constitute a Lie algebra with respect to the bracket
\beq
[J_u,J_{u'}]\op=^{\rm def} J_{[u,u']}. \label{c2}
\eeq
If $Y\to X$ is a vector bundle and $X$ is provided with a non-degenerate
metric $g$, the bracket (\ref{c2}) can be extended to any horizontal
$(n-1)$-forms $\f=\f^\al\om_\al$ on $\Pi$ by the law
\be
[\f,\si]=g_{\al\bt}g^{\m\nu}(\dr^i_\m\f^\al\dr_i\si^\bt-
\dr^i_\m\si^\bt\dr_i\f^\al)\om_\nu.
\ee
Similarly, the bracket of horizontal 1-forms on $\Pi$ is defined.$^{11}$
The bracket (\ref{c2}) looks promising for the current algebra quantization
of the covariant Hamiltonian formalism. We will use this bracket in order to
construct the algebra of supercurrents in the BRST extended Hamiltonian
formalism.

Note that, since the above mentioned Poisson bracket $\{\cH,f\}_V$ is not
preserved under time-dependent transformations, the standard BRST technique,
based on the Lie algebra of constraints, can not be applied in a
straightforward manner to time-dependent mechanics and covariant Hamiltonian
field theory. We generalize the BRST mechanics of 
E.Gozzi and M.Reuter$^{11,12,19}$ in the terms of simple graded manifolds.

\bigskip 

\noindent 
{\bf II. TECHNICAL PRELIMINARIES}
\bigskip

All maps throughout the paper are smooth, while manifolds are real,
finite-dimensional, Hausdorff, second-countable and connected. A base
manifold $X$ is oriented.

Given a fiber bundle $Y\to X$ coordinated by $(x^\la,y^i)$, the 
$s$-order jet manifold $J^sY$ is endowed with the adapted 
coordinates $(x^\la,y^i_\La)$, $0\leq\mid\La\mid\leq s$, where 
$\La$ is a symmetric multi-index $(\la_k...\la_1)$, $\nm\La=k$. 
The repeated jet manifold $J^1J^1Y$ is coordinated by 
$(x^\la,y^i,y^i_\la,\wh y^i_\la,y^i_{\la\m})$. 
There are the canonical morphisms
\ben
&&\la=dx^\la \otimes (\dr_\la + y^i_\la \dr_i):J^1Y\op\hookrightarrow_Y T^*X
\op\ot_Y TY,\label{18}\\ 
&&S_1= (\wh y_\la^i -y_\la^i)dx^\la\ot\dr_i:J^1J^1Y\op\to_Y T^*X\op\ot_Y
VY.\label{cmp18}
\een

Exterior forms $\f$ on a manifold $J^sY$, $s=0,1,\ldots$, are naturally
identified with their pull-backs onto $J^{s+1}Y$. There is the exterior algebra
homomorphism, called the horizontal projection,
\beq
h_0:\f_\la dx^\la+ \f_i^\La dy^i_\La \mapsto \f_\la
dx^\la+\f_i^\La y^i_{\la+\La}dx^\la \label{cmp100}
\eeq
which sends exterior forms on $J^sY$ onto the horizontal forms on
$J^{s+1}Y\to X$,
and vanishes on the contact forms $\th^i_\La=dy^i_\La
-y^i_{\la+\La}dx^\la$.
 Note that the horizontal projection $h_0$ and the
pull-back operation with respect to bundle morphisms over $X$
mutually commute.  Recall also the operators of 
the total derivative 
\be
d_\la =
\dr_\la +y^i_{\la+\La}\dr^\La_i=\dr_\la +y^i_\la\dr_i + y^i_{\la\m}\dr_i^\m
+\cdots,
\ee
and the horizontal differential 
$d_H\f=dx^\la\w d_\la\f$ such that $h_0\circ d=d_H\circ h_0$.

We regard a connection on a fiber bundle $Y\to X$ as a
global section 
\beq
\G=dx^\la\ot(\dr_\la +\G^i_\la\dr_i) \label{cmp91}
\eeq
of the affine jet bundle $\pi^1_0:J^1Y\to Y$.$^{3,20}$
Sections of the underlying vector bundle $T^*X\op\ot_YVY\to Y$ are called
soldering forms. 
Every connection $\G$ on a fiber bundle $Y\to X$
gives rise to the connection
\beq
V\G=dx^\la\ot(\dr_\la +\G^i_\la\dr_i +\dr_j\G^i_\la\dot y^j
\frac{\dr}{\dr\dot y^i}) \label{cmp43}
\eeq
on the fiber bundle $VY\to X$.

\bigskip 

\noindent 
{\bf III. LAGRANGIAN DYNAMICS}
\bigskip

We follow the first variational formula of the
calculus of variations.$^3$ Given a Lagrangian $L$ and its Lepagean
equivalent
$H_L$, this formula provides the canonical decomposition of the Lie derivative
of $L$ along a projectable vector field $u$ on $Y$ in accordance with the
variational problem. We restrict our consideration to
the Poincar\'e--Cartan form
\beq
 H_L=\cL\om +\pi^\la_i\th^i\w\om_\la, \quad \pi^\la_i=\dr^\la_i\cL, \quad
\om_\la=\dr_\la\rfloor\om.
\label{303}
\eeq
In contrast with other Lepagean equivalents,
$H_L$ is a horizontal form on the affine jet bundle $J^1Y\to Y$. Moreover, it
is the Lagrangian counterpart of polysymplectic Hamiltonian forms (see the
relations (\ref{4.9}) and (\ref{2.32}) below). The first variational formula
reads
\beq
\bL_{J^1u}L=
 u_V\rfloor \cE_L + d_Hh_0(u\rfloor H_L), \label{C30} 
\eeq
where $u_V=(u\rfloor\th^i)\dr_i$ and
\beq
\cE_L=
 (\dr_i- d_\la\dr^\la_i)\cL \th^i\w\om: J^2Y\to T^*Y\w(\op\w^nT^*X) \label{305}
\eeq
is the Euler--Lagrange operator associated with $L$. The kernel
of $\cE_L$ defines the Euler--Lagrange equations on
$Y$ given by the coordinate relations
\beq
(\dr_i- d_\la\dr^\la_i)\cL=0. \label{b327} 
\eeq
Solutions of these equations are 
critical sections of the
variational problem for the Lagrangian $L$. 

\begin{rem} 
The first variational formula (\ref{C30}) also
provides the Lagrangian conservation laws.$^{3,21}$ On-shell, we have the
weak identity
\be
 \bL_{J^1u}L\ap d_Hh_0(u\rfloor H_L), 
\ee
and, if $\bL_{J^1u}L=0,$
the weak conservation law 
\beq
0\ap d_Hh_0(u\rfloor H_L)= - d_\la \cT^\la \om \label{K4}
\eeq
of the symmetry current
\beq
\cT =-h_0(u\rfloor H_L)=\cT^\la\om_\la =-[\pi^\la_i(u^\m y^i_\m-u^i
)-u^\la\cL]\om_\la \label{Q30}
\eeq
along the vector field $u$. 
\end{rem}
 
Instead of the variational problem on $Y$ for a Lagrangian $L$,
one can consider that on $J^1Y$ for the  Poincar\'e--Cartan form
$H_L$ (\ref{303}). 
Critical sections $\ol s: X\to J^1Y$ of this variational problem
satisfy the relation
\beq
 \ol s^*(u\rfloor dH_L)=0 \label{C28}
\eeq
for all vertical vector fields $u$ on $J^1Y\to X$. This relation defines the
Cartan equations on $J^1Y$.
We regain these equations in another way.$^3$

Let us consider the above-mentioned Legendre map 
\be
\wh L:J^1Y \op\to_Y \Pi, \qquad p^\la_i\circ\wh L =\pi^\la_i.
\ee
The Legendre bundle $\Pi$ (\ref{00}) is equipped with the canonical
tangent-valued Liouville form $\th$ (\ref{2.4}). Its pull-back on $J^1Y$ by
$\wh L$ is
\be
\th_L =  \wh L^*\th = - \pi_i^\la dy^i\w\om\ot \dr_\la.
\ee
We construct the
reduced Lagrangian 
\beq
\ol L = L - S_1\rfloor\th_L = (\cL + (\wh y_\la^i - y_\la^i)\pi_i^\la)\om 
\label{cmp80}
\eeq
on $J^1J^1Y$ (see the notation (\ref{cmp18})). The
associated Euler--Lagrange operator, called the Euler--Lagrange--Cartan
operator for $L$, reads
\ben
&& \cE_{\ol L} : J^1J^1Y\to T^*J^1Y\w(\op\w^n T^*X), \nonumber \\
&& \cE_{\ol L} = [(\dr_i\cL - \wh d_\la\pi_i^\la 
+ \dr_i\pi_j^\la(\wh y_\la^j - y_\la^j))dy^i + \dr_i^\la\pi_j^\m(\wh
y_\m^j - y_\m^j) dy_\la^i]\w \om, \label{2237} \\
&&\wh d_\la=\dr_\la +\wh y^i_\la\dr_i +y^i_{\la\m}\dr_i^\m. \nonumber
\een
This is the Lagrangian counterpart of the polysymplectic Hamilton operator
(see the relation (\ref{b4.1000}) below). Its kernel
$\Ker\cE_{\ol L}\subset J^1J^1Y$ is given exactly by the Cartan equations
\bea
&& \dr_i^\la\pi_j^\m(\wh y_\m^j - y_\m^j)=0, \label{b336a}\\
&& \dr_i \cL - \wh d_\la\pi_i^\la 
+ (\wh y_\la^j - y_\la^j)\dr_i\pi_j^\la=0. \label{b336b}
\eea
Since $\cE_{\ol L}\mid_{J^2Y}=\cE_L$, the Cartan equations (\ref{b336a}) --
(\ref{b336b}) are equivalent to the Euler--Lagrange equations 
 (\ref{b327}) on integrable sections of
$J^1Y\to X$. These equations are equivalent 
in the case of regular Lagrangians.

With the Poincar\'e--Cartan form $H_L$ (\ref{303}), 
we have the Legendre morphism
\be
\wh H_L: J^1Y\op\to_Y Z_Y, \qquad 
(p^\m_i, p)\circ\wh H_L =(\pi^\m_i, \cL-\pi^\m_i y^i_\m ), 
\ee
where the fiber bundle $Z_Y$ (\ref{N41}) is endowed with holonomic
coordinates $(x^\la,y^i,p^\la_i,p)$. 
It is readily observed that
\beq
\wh L=\pi_{Z\Pi}\circ \wh H_L. \label{LZP}
\eeq
Owing to the monomorphism $Z_Y\hookrightarrow \op\w^nT^*Y$, the bundle
$Z_Y$ is equipped with the pull-back
\beq
\Xi_Y= p\om + p^\la_i dy^i\w\om_\la \label{N43}
\eeq
of the canonical form $\Theta$ on 
$\op\w^nT^*Y$ whose exterior differential $d\Theta$ is the $n$-multisymplectic
form in the sense of Martin.$^{22}$ 

Let $Z_L=\wh H_L(J^1Y)$  be an imbedded subbundle
$i_L:Z_L\hookrightarrow Z_Y$ of $Z_Y\to Y$.
It is provided with the pull-back 
 De Donder form $i^*_L\Xi_Y$. We have 
\beq
H_L=\wh H_L^*\Xi_L=\wh H_L^*(i_L^*\Xi_Y).  \label{cmp14}
\eeq
By analogy with the Cartan equations
(\ref{C28}), the  Hamilton--De Donder equations for sections $\ol r$ of 
$Z_L\to X$ are written as
\beq
\ol r^*(u\rfloor d\Xi_L)=0 \label{N46}
\eeq
where $u$ is an arbitrary vertical vector field on
$Z_L\to X$. To obtain an explicit form of these equations, one should
substitute solutions
$(y^i_\la$, $\cL)$
 of the equations
\beq
p^\la_i=\pi^\la_i, \qquad
p=\cL -\pi^\la_i y^i_\la  \label{N60}
\eeq
in the Cartan equations.
However, if a Lagrangian $L$ is
degenerate, the equations (\ref{N60}) may admit different solutions
or no solution at all. Something more is said in the following theorem.$^1$

\begin{theo}\label{ddd}  Let the Legendre morphism
$\wh H_L:J^1Y\to Z_L$ be a submersion. Then a section $\ol s$ of $J^1Y\to X$
is a solution of the Cartan equations (\ref{C28}) if and only if $\wh
H_L\circ\ol s$ is a solution of the Hamilton--De Donder equations
(\ref{N46}), i.e., Cartan and Hamilton--De Donder equations are
quasi-equivalent.
\end{theo}

\bigskip 

\noindent 
{\bf IV. COVARIANT HAMILTONIAN DYNAMICS}
\bigskip

Given a fiber bundle $Y\to X$, let $\Pi$ be the
Legendre bundle (\ref{00}). Holonomic coordinates
$(x^\la ,y^i,p^\la_i)$ on $\Pi$  are compatible with its composite fibration
\be
\pi_{\Pi X}=\pi\circ\pi_{\Pi Y}:\Pi\to Y\to X.
\ee
We have the canonical bundle monomorphism 
\beq
\th =-p^\la_idy^i\w\om\otimes\dr_\la
:\Pi\op\hookrightarrow_Y\op\w^{n+1}T^*Y\op\otimes_Y TX, \label{2.4}
\eeq
called the tangent-valued Liouville form on $\Pi$.
It should be emphasized that the exterior differential $d$ can not be applied
to the tangent-valued form (\ref{2.4}). At the same time, there is a unique
$TX$-valued $(n+2)$-form
$\Om_Y$ on
$\Pi$ such that the relation
\be
\Om_Y\rfloor\f =-d(\th\rfloor\f)
\ee
holds for any exterior 1-form $\f$ on $X$.$^3$ 
This form, called the polysymplectic form, is given by the coordinate
expression
\beq
\Om_Y =dp_i^\la\w dy^i\w \om\ot\dr_\la. \label{406}
\eeq

As was mentioned above,  every section $h$ of the fiber bundle (\ref{b418'})
defines the pull-back (\ref{b418}) of the canonical form $\Xi_Y$ (\ref{N43}),
called a Hamiltonian form on the Legendre bundle $\Pi$.

\begin{prop}\label{sphamf} 
Hamiltonian forms on $\Pi$ constitute a non-empty affine space modelled
over the linear space of horizontal densities $\wt H=\wt{\cH}\om$
on $\Pi\to X$. 
\end{prop}

\begin{proof}
This is an immediate consequence of the fact that
(\ref{b418'}) is an affine bundle modelled over the pull-back vector bundle
$\Pi\op\times_X\op\w^nT^*X\to \Pi$.
\end{proof}

\begin{lem}$^{2,4}$ \label{cmp2}  Every connection
$\G$ (\ref{cmp91}) on $Y\to X$ yields the splitting 
\be
h_\G: \ol dy^i \mapsto dy^i- \G^i_\la dx^\la
\ee
of the exact sequence (\ref{b417}) and, as a consequence, defines the
Hamiltonian form 
\ben
H_\G =h_\G^*\Xi_Y =p^\la_i dy^i\w\om_\la -p^\la_i\G^i_\la\om. \label{3.6}
\een
\end{lem}

 Proposition \ref{sphamf} and Lemma \ref{cmp2} lead to the
following

\begin{cor}\label{hams} 
Given a connection $\G$ on $Y\to X$, every 
Hamiltonian form $H$ admits the decomposition
\beq
H=H_\G -\wt H_\G =p^\la_idy^i\w\om_\la
-p^\la_i\G^i_\la\om-\wt{\cH}_\G\om. \label{4.7}
\eeq
\end{cor}

\begin{rem}
The physical meaning of the splitting (\ref{4.7}) is illustrated  by
the fact that, in the case of $X=\bR$, $\wt\cH_\G$ is exactly the energy of
a mechanical system with respect to the reference frame $\G$.$^{10,11}$ 
\end{rem}

We will mean by a Hamiltonian map any bundle morphism
\beq
\Phi=dx^\la\otimes(\dr_\la +\Phi^i_\la\dr_i):\Pi\op\to_Y J^1Y.
\label{2.7}
\eeq
In particular, let $\G$ be a connection on $Y\to X$. Then, the
composition 
\beq
\wh\G=\G\circ\pi_{\Pi Y}=dx^\la\otimes (\dr_\la +\G^i_\la\dr_i):\Pi\to Y\to
J^1Y, \label{b420}
\eeq
is a Hamiltonian map. Conversely,
every Hamiltonian map $\Phi$ yields
the associated connection $\G_\Phi =\Phi\circ\wh 0$
on $Y\to X$, where $\wh 0$ is the global zero section of the
Legendre bundle $\Pi\to Y$. In particular, we have
$\G_{\wh\G}=\G$. The following two facts will be used in the sequel.

\begin{prop}$^3$ 
Every Hamiltonian form $H$ (\ref{b418}) yields the Hamiltonian
map $\wh H$ such that
\beq
 y_\la^i\circ\wh H=\dr^i_\la\cH. \label{415}
\eeq
\end{prop}

\begin{prop} Every Hamiltonian map 
(\ref{2.7}) defines the Hamiltonian form
\be
H_\Phi=\Phi\rfloor\th =p^\la_idy^i\w\om_\la -p^\la_i\Phi^i_\la\om.
\ee
\end{prop}

\begin{proof}
Given an arbitrary connection $\G$ on the fiber bundle $Y\to X$, the
corresponding Hamiltonian map (\ref{b420}) defines the form $\wh\G\rfloor\th$
which is exactly the Hamiltonian form $H_\G$ (\ref{3.6}). Since $\Phi-\wh\G$
is a $VY$-valued basic 1-form on $\Pi\to X$, $H_\Phi-H_\G$ is a
horizontal density on $\Pi$. Then the result follows from Proposition
\ref{sphamf}. Note that $H=H_{\wh H}$ iff $H=H_\G$ (\ref{3.6}).
\end{proof} 

Let $J^1\Pi$ be the first order jet
manifold of 
$\Pi\to X$. It is equipped with the adapted coordinates
$( x^\la ,y^i,p^\la_i,y^i_\m,p^\la_{\m i})$ such that
$y^i_\m\circ J^1\pi_{\Pi Y}=y^i_\m$.
A connection 
\beq
\g =dx^\la\otimes(\dr_\la +\g^i_\la\dr_i
+\g^\m_{\la i}\dr^i_\m) \label{cmp33}
\eeq
on $\Pi\to X$ is called a Hamiltonian
connection if the exterior form $\g\rfloor\Om_Y$
is closed. 
A Hamiltonian connection $\g$ is said to be associated with
a Hamiltonian form $H$ if it obeys the condition
\ben
&& \g\rfloor\Om_Y= dH, \label{cmp3}\\
&& \g^i_\la =\dr^i_\la\cH, \qquad 
\g^\la_{\la i}=-\dr_i\cH. \label{3.10}
\een

\begin{theo}\label{loch} 
For every Hamiltonian connection $\g$, there exists a local Hamiltonian form
$H$ on a neighbourhood of any point $q\in\Pi$ such that the equation
(\ref{cmp3}) holds.
\end{theo}

\begin{proof} If $\g\rfloor\Om_Y$ is closed, there is a
contractible neighbourhood
$U$ of a point $q\in\Pi$ which belongs to a holonomic coordinate chart
$(x^\la,y^i,p^\la_i)$ and where the local form $\g\rfloor\Om_Y$ is exact.
We have
\beq
\g\rfloor\Om_Y =dH=dp^\la_i\w dy^i\w\om_\la -
(\g^i_\la dp^\la_i -\g^\la_{\la i}dy^i)\w\om \label{cmp4}
\eeq
on $U$. It is readily observed that the second term in the right-hand side of
this equality is also an exact form on $U$. By virtue of the relative 
Poincar\'e lemma, it can be brought into the form $d\cH\w\om$ where $\cH$ is a
local function on $U$. Then the form $H$ in the expression
(\ref{cmp4}) reads
\be
H=p^\la_idy^i\w\om_\la -\cH\om.
\ee
Using Corollary \ref{hams}, one can easily show that this is a 
Hamiltonian form on $U$.
\end{proof}

\begin{theo}
Every Hamiltonian form has an associated
Hamiltonian connection.
\end{theo}

\begin{proof}
Given a Hamiltonian form $H$, let us consider the first
order differential operator 
\ben
&& \cE_H :J^1\Pi\to T^*\Pi\w(\op\w^n T^*X),\nonumber \\
&& \cE_H=dH-\la\rfloor\Om_Y=[(y^i_\la-\dr^i_\la\cH) dp^\la_i
-(p^\la_{\la i}+\dr_i\cH) dy^i]\w\om, \label{3.9}
\een 
on $\Pi$ where $\la$ is the canonical monomorphism (\ref{18}). It
is called the Hamilton operator associated with $H$. A glance at the expression
(\ref{3.9}) shows that this operator is an affine morphism over
$\Pi$ of constant rank. It follows that its kernel is an affine
closed imbedded subbundle  of the jet bundle $J^1\Pi\to\Pi$. This subbundle
has a global section $\g$ which is a connection on $\Pi\to X$. This connection
obeys the equation (\ref{cmp3}).
\end{proof}

It should be emphasized that, if $n>1$, there is a set of 
Hamiltonian connections associated with the same Hamiltonian form
$H$. They differ from each other in soldering forms $\si$ on $\Pi\to
X$ which obey the equation $\si\rfloor\Om_Y=0$.

\begin{prop}
Every  Hamiltonian connection $\g$ associated with a Hamiltonian form $H$
satisfies the relation
\beq
J^1\pi_{\Pi Y}\circ \g= \wh H. \label{4.109} 
\eeq
\end{prop}

\begin{proof}
The proof is based on the expressions (\ref{415}) and (\ref{3.10}).
\end{proof}

Being a closed subbundle of the jet
bundle
$J^1\Pi\to X$, the kernel of the Hamilton operator $\cE_H$
(\ref{3.9}) defines first order Hamilton equations (\ref{b4100a}) --
(\ref{b4100b})  on the Legendre bundle
$\Pi$. Every integral section 
$J^1r=\g\circ r$
of a Hamiltonian connection $\g$ associated with a Hamiltonian form $H$ is
obviously a solution of the Hamilton equations (\ref{b4100a}) --
(\ref{b4100b}). Conversely, a solution of the Hamilton equations
(\ref{b4100a}) -- (\ref{b4100b}) is
 a section $r$ of $\Pi\to X$ such that
its jet prolongation $J^1r$ lives in $\Ker\cE_H$.
If  $r:X\to\Pi$
is a global solution, there exists an extension of the local section
$J^1r: r(X) \to J^1\Pi$
to a Hamiltonian connection which has $r$ as an integral
section. Substituting $J^1r$ in (\ref{4.109}), we obtain the equality
\beq
J^1(\pi_{\Pi Y}\circ r)= \wh H\circ r, \label{N10}
\eeq
which is the coordinate-free form of the Hamilton equations (\ref{b4100a}).
Nevertheless, it may happen that the
Hamilton equations 
have no solution through a given point $q\in \Pi$.

\begin{rem}
The Hamilton equations can be introduced
without appealing to the Hamilton operator. 
As was for the Cartan equations (\ref{C28}),
they are equivalent to the condition
\beq
r^*(u\rfloor dH)= 0 \label{N7}
\eeq
for any vertical vector field $u$ on $\Pi\to X$.
\end{rem}

\bigskip 

\noindent 
{\bf V. LAGRANGIAN AND HAMILTONIAN DEGENERATE SYSTEMS}
\bigskip

Let us study the relations between 
Hamilton and Euler--Lagrange equations when
a Lagrangian is degenerate.
Their main peculiarity lies in the fact that
there is a set of 
Hamiltonian forms associated with the same degenerate Lagrangian.

\begin{rem} \label{cmp7} 
Let us recall the coordinate expressions
\ben
&& (y^i_\m,\wh y^i_\la,y^i_{\la\m})\circ
J^1\wh H=(\dr^i_\m\cH,y^i_\la,d_\la\dr^i_\m\cH),\label{cmp5}\\
&&(p^\la_i, y^i_\m,p^\la_{\m i})\circ J^1\wh L=
(\pi^\la_i,\wh y^i_\m, \wh d_\m\pi^\la_i). \label{cmp6}
\een
In particular, if $\g$ is a Hamiltonian connection for a
Hamiltonian form $H$, we obtain from  (\ref{N10}) and (\ref{cmp5}) that
the composition $J^1\wh H\circ\g$ takes its
values 
into the sesquiholonomic jet bundle $\wh J^2Y$.
\end{rem}

A Hamiltonian form $H$ is said to be associated
with a Lagrangian $L$ if $H$ satisfies the relations
\bea
&&\wh L\circ\wh H\circ \wh L=\wh L,\label{2.30a} \\
&&H=H_{\wh H}+\wh H^*L. \label{2.30b}
\eea
A glance at the relation (\ref{2.30a}) shows that $\wh
L\circ\wh H$ is the projector
\beq
p^\m_i(q)=\dr^\m_i\cL (x^\m,y^i,\dr^j_\la\cH(q)), \qquad q\in N_L,
\label{b481'}
\eeq
from $\Pi$ onto the Lagrangian constraint space $N_L=\wh L( J^1Y)$.
Accordingly,  $\wh H\circ\wh L$ is the projector from $J^1Y$ onto $\wh
H(N_L)$. 

\begin{lem} \label{cmp110}
Any Hamiltonian form $H$ associated with a Lagrangian $L$ obeys
the relation
\beq
H\mid_{N_L}=\wh H^*H_L\mid_{N_L}, \label{4.9}
\eeq
where $H_L$ is the Poincar\'e--Cartan form (\ref{303}).
\end{lem}

\begin{proof}
The relation (\ref{2.30b}) takes the coordinate form
\beq
\cH=p^\m_i\dr^i_\m\cH-\cL(x^\m,y^i,\dr^j_\la\cH). \label{b481}
\eeq
Substituting (\ref{b481'}) and (\ref{b481}) in (\ref{b418}), we
obtain the relation (\ref{4.9}).
\end{proof}

Something more can be said in the case of semiregular Lagrangians.
A Lagrangian $L$ is called semiregular if the pre-image $\wh L^{-1}(q)$ of 
any point $q\in N_L$ is a connected submanifold of $J^1Y$. The following fact
will be used in the sequel.

\begin{lem} \label{3.22} 
The 
Poincar\'e--Cartan form $H_L$ for a 
semiregular Lagrangian $L$ is constant on the connected
pre-image $\wh L^{-1}(q)$ of any point  $q\in N_L$.
\end{lem}

\begin{proof}
Let $u$ be a vertical vector field on the affine jet bundle $J^1Y\to
Y$ which takes its values into the kernel of the
tangent map $T\wh L$ to $\wh L$. Then $\bL_u H_L=0$. 
\end{proof}

An immediate consequence of this fact is the following assertion.

\begin{prop} \label{3.22'}  All Hamiltonian forms associated
with a semiregular Lagrangian $L$ coincide with each
other on the Lagrangian constraint space $N_L$,
and the Poincar\'e--Cartan form $H_L$ (\ref{303})
for $L$ is the pull-back
\ben
&& H_L=\wh L^*H, \label{2.32}\\
&& (\pi^\la_iy^i_\la-\cL)\om=\cH(x^\m,y^j,\pi^\m_j)\om,\nonumber
\een
of any such a Hamiltonian form $H$.
\end{prop}

\begin{proof}
Given a
vector $v\in T_q\Pi$, the value
$T\wh H(v)\rfloor H_L(\wh H(q))$ is the same for all Hamiltonian maps
$\wh H$ satisfying the relation (\ref{2.30a}). Then the results follow from
the relation (\ref{4.9}).
\end{proof}

Proposition \ref{3.22'} enables us to connect Euler--Lagrange and Cartan
equations for a semiregular Lagrangian $L$ with the Hamilton
equations for Hamiltonian forms associated with $L$.

\begin{theo}\label{3.23} 
Let a section $r$ of $\Pi\to X$
be a  solution of the Hamilton equations (\ref{b4100a}) -- (\ref{b4100b})
for a Hamiltonian form $H$ associated with a semiregular
Lagrangian $L$. If $r$ lives in the constraint space $N_L$, the
section $s=\pi_{\Pi Y}\circ r$
of $Y\to X$ satisfies the Euler--Lagrange
equations (\ref{b327}), while $\ol s=\wh H\circ r$ obeys the Cartan equations
(\ref{b336a}) -- (\ref{b336b}). 
\end{theo}

\begin{proof}
 Acting by the exterior differential on the relation
(\ref{2.32}), we obtain the relation
\beq
(y^i_\la-\dr^i_\la\cH\circ\wh L)d\pi^\la_i\w\om-(\dr_i\cL+ \dr_i(\cH\circ\wh
L))dy^i\w\om=0 \label{b4160}
\eeq
which is equivalent to the system of equalities
\be
&&\dr^\la_i\pi^\m_j(y^j_\m -\dr^j_\m\cH\circ\wh L)=0, \\
&& \dr_i\pi^\m_j(y^j_\m -\dr^j_\m\cH\circ\wh L) -(\dr_i\cL +(\dr_i\cH)\circ\wh
L)=0.
\ee
Using these equalities and the expression (\ref{cmp6}), one can easily see that
\beq 
\cE_{\ol L}=(J^1\wh L)^*\cE_H, \label{b4.1000}
\eeq
where $\cE_{\ol L}$ is the Euler--Lagrange--Cartan operator (\ref{2237}). 
Let $r$ be a section of $\Pi\to X$ which lives in the Lagrangian constraint
space $N_L$, and
$\ol s=\wh H\circ r$. Then we have
\be
r=\wh L\circ \ol s, \qquad J^1r=J^1\wh L\circ J^1\ol s.
\ee 
If $r$ is a solution of the Hamilton equations, the exterior form $\cE_H$
vanishes on $J^1r(X)$. Hence, the  pull-back form $\cE_{\ol
L}=(J^1\wh L)^*\cE_H$ vanishes on $J^1\ol s(X)$. It follows that
$\ol s$  obeys the Cartan equations (\ref{b336a}) -- (\ref{b336b}). We obtain
from the equality (\ref{N10}) that $\ol s= J^1s$, $s=\pi_{\Pi Y}\circ r$.
Hence, $s$ is a solution of the Euler--Lagrange equations.
\end{proof}

The same result can be obtained from the relation
\beq
\ol L=(J^1\wh L)^*L_H \label{cmp83}
\eeq
where $\ol L$ is the Lagrangian (\ref{cmp80}) on $J^1J^1Y$ and $L_H$ is the
Lagrangian (\ref{Q3}) on $J^1\Pi$.

\begin{theo}\label{3.24} Given a semiregular Lagrangian $L$,
let a section $\ol s$ of the jet bundle
$J^1Y\to X$ be a solution of the
Cartan equations (\ref{b336a}) -- (\ref{b336b}).
Let $H$ be a Hamiltonian form associated with $L$,  and let $H$ satisfy the
relation
\beq
\wh H\circ \wh L\circ \ol s=J^1(\pi^1_0\circ\ol s).\label{2.36}
\eeq
Then, the section $r=\wh L\circ \ol s$
of the Legendre bundle $\Pi\to X$ is a solution of the
Hamilton equations (\ref{b4100a}) -- (\ref{b4100b}) for $H$.
\end{theo}

\begin{proof} The Hamilton equations (\ref{b4100a}) hold by
virtue of the condition (\ref{2.36}). Substituting $\wh L\circ \ol s$
in the Hamilton equations (\ref{b4100b}) and
using the relations
(\ref{b4160}) and (\ref{2.36}), we come to the  Cartan equations
(\ref{b336b}) for $\ol s$ as follows: 
\be
&& \wh d_\la \pi^\la_i\circ\ol s +(\dr_i\cH)\circ\wh L\circ \ol s=
\wh d_\la \pi^\la_i\circ\ol s + (\ol s^j_\m
-\dr^j_\m\cH\circ\wh L\circ \ol s)\dr_i\pi^\m_j\circ\ol s -\dr_i\cL\circ\ol s=
\\ && \qquad \wh d_\la \pi^\la_i\circ\ol s- (\dr_\m \ol s^j -\ol
s^j_\m)\dr_i\pi^\m_j\circ\ol s -\dr_i\cL\circ\ol s=0.
\ee
\end{proof}

\begin{rem} \label{cmp9} 
Since $\wh H\circ \wh L$ in Theorem (\ref{3.24}) is a projection operator,
the condition (\ref{2.36}) implies that the solution $\ol s$ of the Cartan
equations is actually an integrable section $\ol s=J^1s$ where $s$ is a
solution of the Euler--Lagrange equations.  
Theorems \ref{3.23} and \ref{3.24} show that, if a solution of the Cartan
equations provides a solution of the covariant Hamilton equations, it is
necessarily a solution of the Euler--Lagrange equations. In fact, the
relation (\ref{b4.1000}) gives more than it is needed for proving Theorem
\ref{3.23}. Using this relation, one can justify that, if $\g$ is a Hamiltonian
connection for a Hamiltonian form $H$ associated with a semiregular Lagrangian
$L$, then the composition $J^1\wh H\circ\g\circ\wh L$ takes its values in
$\Ker \cE_{\ol L}\cap\wh{J^2Y}$ (see Remark \ref{cmp7}), i.e., this is a local
sesquiholonomic Lagrangian connection on
$\wh H(N_L)$.$^3$  A converse of this assertion, however, fails to
be true in the case of semiregular Lagrangians. 
Let a Lagrangian $L$ be
hyperregular, i.e., the Legendre map $\wh L$
is a diffeomorphism. Then $\wh L^{-1}$ is a Hamiltonian map, and there is a
unique Hamiltonian form 
\beq
H=H_{\wh L^{-1}}+\wh L^{-1*}L \label{cmp8}
\eeq
associated with $L$. In this case, both the relation (\ref{b4.1000}) and the
converse one 
\be
\cE_H=(J^1\wh H)^*\cE_{\ol L}
\ee
hold. It follows that the Euler--Lagrange equations for $L$ and the Hamilton
equations for $H$ (\ref{cmp8}) are equivalent.
\end{rem}

We will say that a set of Hamiltonian forms
$H$  associated with a semiregular Lagrangian $L$ is
complete if, for each
solution
$s$ of the Euler-Lagrange equations, there exists a solution
$r$ of the Hamilton equations for a Hamiltonian form $H$ from this 
set such that $s=\pi_{\Pi Y}\circ r$.
By virtue of Theorem \ref{3.24} and Remark \ref{cmp9}, a set
of associated Hamiltonian forms
is complete if, for every solution $s$ on $X$ of the Euler-Lagrange
equations for $L$, there is a Hamiltonian form $H$ from this
set which fulfills the relation 
\beq
\wh H\circ \wh L\circ J^1s=J^1s. \label{2.36'}
\eeq

As for the existence of complete sets of associated Hamiltonian forms,
we refer to the following theorem.  A Lagrangian $L$ is said to be almost
regular if (i) $L$ is semiregular, 
(ii) the Lagrangian constraint space $N_L$  is a closed
imbedded subbundle $i_N:N_L\hookrightarrow\Pi$ of the Legendre bundle $\Pi\to
Y$ and (iii) the Legendre map 
\beq
\wh L:J^1Y\to N_L \label{cmp12}
\eeq
is a submersion, i.e., 
a fibred manifold.

\begin{prop}$^{2,23}$  Let $L$ be an almost regular
Lagrangian. On an open neighbourhood in $\Pi$ of each point $q\in N_L$, there
exist local Hamiltonian forms associated with $L$ which constitute  a
complete set.
\end{prop}

In the case of an almost regular Lagrangian $L$, we can say something more on
the relations between Lagrangian and Hamiltonian systems as follows.
Let us assume that the fibred manifold (\ref{cmp12}) admits a
global section
$\Psi$. Let us consider the pull-back 
\beq
H_N=\Psi^*H_L, \label{b4300}
\eeq
called the constrained Hamiltonian form.
By virtue of Lemma
\ref{3.22}, it is uniquely defined for all sections of the fibred manifold
$J^1Y\to N_L$, and $H_L=\wh L^*H_N$.
For sections
$r$ of the fiber bundle $N_L\to X$, we can write the 
constrained Hamilton equations
\beq
r^*(u_N\rfloor dH_N) =0 \label{N44}
\eeq
where $u_N$ is an arbitrary vertical vector field on $N_L\to X$. These
 equations possess the following important properties.

\begin{theo} \label{cmp22} For any Hamiltonian form $H$ 
associated with an almost regular Lagrangian $L$, every solution $r$
of the Hamilton equations which lives in the Lagrangian constraint space $N_L$
is a solution of the constrained Hamilton equations (\ref{N44}).
\end{theo}

\begin{proof} Such a Hamiltonian form $H$ defines 
the global section $\Psi=\wh H\circ i_N$ of the fibred manifold (\ref{cmp12}).
Due to the relation (\ref{2.32}), $H_N=i^*_NH$ and the constrained Hamilton
equations can be written as
\beq
r^*(u_N\rfloor di^*_NH)=r^*(u_N\rfloor dH\mid_{N_L}) =0. \label{N44'}
\eeq
Note that they differ from the Hamilton equations 
(\ref{N7}) restricted to $N_L$ which read
\beq
r^*(u\rfloor dH\mid_{N_L}) =0, \label{cmp10}
\eeq
where $r$ is a section of $N_L\to X$ and $u$ is an arbitrary vertical vector
field on $\Pi\to X$. A solution $r$ of the equations
(\ref{cmp10}) satisfies obviously the weaker condition
(\ref{N44'}).
\end{proof}

\begin{theo}\label{3.02} 
The constrained Hamilton equations (\ref{N44}) are equivalent to the
Hamilton--De Donder equations (\ref{N46}).
\end{theo}

\begin{proof} 
In accordance with the relation (\ref{LZP}), the projection $\pi_{Z\Pi}$
(\ref{b418'}) yields a surjection of $Z_L$ onto $N_L$. Given a section 
$\Psi$ of the fibred manifold (\ref{cmp12}), we have the morphism
\be
\wh H_L\circ \Psi: N_L\to Z_L.
\ee
By virtue of Lemma (\ref{3.22}), this is a surjection  such that 
\be
\pi_{Z\Pi}\circ\wh H_L\circ \Psi=\id_{N_L}.
\ee
Hence, $\wh H_L\circ \Psi$ is a bundle isomorphism over $Y$ which is
independent of the choice of a global section $\Psi$. Combining (\ref{cmp14})
and (\ref{b4300}) gives
\be
H_N=(\wh H_L\circ \Psi)^*\Xi_L
\ee
that leads to the desired equivalence.
\end{proof}

The above proof gives more. Namely, since $Z_L$ and $N_L$ are isomorphic,
the Legendre morphism $H_L$ fulfills
the conditions of Theorem \ref{ddd}. Then combining Theorem \ref{ddd} and
Theorem \ref{3.02}, we obtain the following theorem.

\begin{theo}\label{3.01}  Let $L$ be an almost regular Lagrangian
such that the fibred manifold (\ref{cmp12}) has a global section. A section
$\ol s$ of the jet bundle
$J^1Y\to X$ is a solution of the Cartan equations (\ref{C28}) iff
$\wh L\circ
\ol s$ is a solution of  the constrained Hamilton equations (\ref{N44}).
\end{theo}

Theorem \ref{3.01} is also a corollary of the following Lemma \ref{cmp84}.
The constrained Hamiltonian form $H_N$ (\ref{b4300}) defines the
constrained Lagrangian 
\beq
L_N=h_0(H_N)=(J^1i_N)^*L_H \label{cmp81}
\eeq
on the jet manifold $J^1N_L$ of the fiber bundle $N_L\to X$.

\begin{lem} \label{cmp84}
There are the relations
\beq
\ol L=(J^1\wh L)^*L_N, \qquad L_N=(J^1\Psi)^*\ol L, \label{cmp85}
\eeq
where $\ol L$ is the Lagrangian (\ref{cmp80}).
\end{lem}

\begin{proof}
The first of the relations (\ref{cmp85}) is an immediate consequence of the
relation (\ref{cmp83}). The latter follows from the expression (\ref{cmp5})
and the relation (\ref{b481'}) if we put $\Psi=\wh H\circ i_N$ for some
Hamiltonian form $H$ associated with the almost regular Lagrangian $L$.
\end{proof}

The Euler--Lagrange equation for the constrained Lagrangian $L_N$ (\ref{cmp81})
are equivalent to the constrained Hamilton equations and, by virtue of Lemma
\ref{cmp84}, are quasi-equivalent to the Cartan equations. At the same
time, Cartan equations of degenerate Lagrangian systems contain an additional
freedom in comparison with the restricted Hamilton equations (see the next
Section).

The correspondence between Lagrangian and covariant Hamiltonian
dynamics of classical fields can be extended to symmetry currents and
conservation laws as follows.$^{3,21}$
Given a projectable vector
field $u$ on a fiber bundle $Y\to X$ and its lift
\beq
\wt u=u^\m\dr_\m + u^i\dr_i +( - \dr_i u^j p^\la_j -\dr_\m u^\m p^\la_i
+\dr_\m u^\la p^\m_i)\dr^i_\la \label{cmp90}
\eeq
onto the Legendre bundle $\Pi$, we obtain
\beq
\bL_{\wt u}H= \bL_{J^1\wt u}L_H, \label{b4180}
\eeq
i.e., the Hamiltonian form $H$ (\ref{b418}) and the
Lagrangian  $L_H$ (\ref{Q3}) have the same symmetries. If the Lie derivatives
(\ref{b4180}) vanish, the corresponding symmetry current 
\beq
J_u = h_0(\wt u\rfloor H)\mid_{y^i_\m=\dr^i_\m\cH} \label{991}
\eeq
is conserved on-shell. In particular, if $u$ is a vertical vector field, we
have
\beq
J_u =\wt u\rfloor H=u\rfloor H. \label{cmp129}
\eeq

\begin{prop}  Let a 
Hamiltonian form $H$ be associated
with a semiregular Lagrangian $L$. Let $r$ be a solution
of the Hamilton equations for $H$ which lives
in the Lagrangian constraint space $N_L$. Let $s=\pi_{\Pi Y}\circ r$ be the
corresponding solution  of the Euler--Lagrange equations for
$L$ so that the relation (\ref{2.36'}) holds. Then, for any 
projectable vector field $u$ on the fiber bundle $Y\to X$, we have 
\be
J_u(r)=\cT_u(\wh H\circ r),\qquad
J_u(\wh L\circ J^1s) =\cT_u(s), 
\ee
where $\cT$ is the current (\ref{Q30}) on $J^1Y$  and $J$ is the
current (\ref{991}) on $\Pi$.
\end{prop}

It follows that the constrained Hamilton equations have symmetries of the
Euler--Lagrange equations. At the same time, given a Hamiltonian form $H$
associated with a degenerate Lagrangian $L$, the Lagrangian $L_H$ (\ref{Q3})
contains gauge fixing terms in comparison with $L$
and the Lagrangian $L_N$ (\ref{cmp81}) (see the next Section).
\bigskip 

\noindent 
{\bf VI. QUADRATIC DEGENERATE SYSTEMS}
\bigskip

This Section is devoted to the physically important case of almost regular
quadratic Lagrangians. 
Given a fiber bundle $Y\to X$,
let us consider a  quadratic Lagrangian $L$ which has the coordinate
 expression
\beq
\cL=\frac12 a^{\la\m}_{ij} y^i_\la y^j_\m +
b^\la_i y^i_\la + c, \label{N12}
\eeq
where $a$, $b$ and $c$ are local functions on $Y$. This property is
coordinate-independent due to the affine transformation law of the coordinates
$y^i_\la$. The associated Legendre map 
\beq
p^\la_i\circ\wh L= a^{\la\m}_{ij} y^j_\m +b^\la_i \label{N13}
\eeq
is an affine morphism over $Y$. It defines the corresponding linear
morphism
\beq
\ol L: T^*X\op\otimes_YVY\op\to_Y\Pi,\qquad p^\la_i\circ\ol
L=a^{\la\m}_{ij}\ol y^j_\m, \label{N13'}
\eeq
where $\ol y^j_\mu$ are  bundle coordinates on the vector
bundle $T^*X\op\ot_Y VY$.

\begin{lem} The Lagrangian (\ref{N12}) is semiregular.
\end{lem}

\begin{proof} Solutions $y^i_\m$ of the pointwise linear algebraic
equations (\ref{N13}) form  an affine space
modelled over the linear space of solutions of the linear
algebraic equations $a^{\la\m}_{ij} \ol y^j_\m=0$.
At each point of $N_L$, these spaces are obviously connected.
\end{proof}

Let the Lagrangian $L$ (\ref{N12}) be almost regular, i.e.,
the matrix function $a^{\la\m}_{ij}$ is of constant rank. Then
the Lagrangian constraint space $N_L$ 
(\ref{N13}) is an affine subbundle of the Legendre bundle $\Pi\to Y$, modelled
over the vector subbundle $\ol N_L$ (\ref{N13'}) of  $\Pi\to
Y$. 
Hence, $N_L\to Y$ has a global section. For the sake of simplicity, let us
assume that it is the canonical
zero section $\wh 0(Y)$ of $\Pi\to Y$. Then $\ol N_L=N_L$.
Accordingly, the kernel
of the Legendre map (\ref{N13})  is an affine
subbundle of the affine jet bundle $J^1Y\to Y$, modelled over the kernel of
the linear morphism $\ol L$ (\ref{N13'}). Then there exists a connection 
\ben
&&\G: Y\to \Ker\wh L\subset J^1Y, \label{N16}\\
&& a^{\la\m}_{ij}\G^j_\m + b^\la_i =0, \label{250}
\een
on $Y\to X$.
Connections (\ref{N16}) constitute an affine space modelled over the linear
space of soldering forms $\f$ on $Y\to X$ satisfying the conditions
\beq
a^{\la\m}_{ij}\f^j_\m =0 \label{cmp21}
\eeq
and, as a consequence, the conditions $\f^i_\la b^\la_i=0$.
If the Lagrangian (\ref{N12}) is regular, the
connection (\ref{N16}) is unique.

\begin{lem}\label{04.2}  There exists a linear bundle
map
\beq
\si: \Pi\op\to_Y T^*X\op\otimes_YVY, \qquad
\ol y^i_\la\circ\si =\si^{ij}_{\la\m}p^\m_j, \label{N17}
\eeq
such that $\ol L\circ\si\circ i_N= i_N$.
\end{lem}

\begin{proof} 
The map (\ref{N17}) is a solution of the algebraic equations
\beq
a^{\la\mu}_{ij}\si^{jk}_{\mu\al}a^{\al\nu}_{kb}=a^{\la\nu}_{ib}.
\label{N45}
\eeq
After pointwise diagonalization, the matrix
$a$ has some non-vanishing components $a^{AA}$, $A\in I$. Then a
solution of the equations (\ref{N45}) takes the form 
\be
\si_{AA}=(a^{AA})^{-1}, \quad \si_{AA'}=0, \quad A\neq A',\quad A,A'\in
I, 
\ee
while the remaining components $\si_{BC}$, $B\not\in I$, are arbitrary.
 In particular, there is a solution with
\beq
 \si_{BC}=0, \qquad B\not\in I.
\label{b4120}
\eeq
It satisfies the particular relation
\beq
\si=\si\circ\ol L\circ\si. \label{N21}
\eeq
Further on, we will take $\si$ to be the solution (\ref{b4120}).
If the Lagrangian (\ref{N12}) is regular, the linear map
(\ref{N17}) is uniquely determined  by the equations (\ref{N45}).
\end{proof}

The following theorem is the key point of our consideration.

\begin{theo}
There are the splittings
\bea
&& J^1Y=\cS(J^1Y)\op\oplus_Y \cF(J^1Y)=\Ker\wh L\op\oplus_Y{\rm Im}(\si\circ
\wh L), \label{N18} \\
&& y^i_\la=\cS^i_\la+\cF^i_\la= [y^i_\la
-\si^{ik}_{\la\al} (a^{\al\m}_{kj}y^j_\m + b^\al_k)]+
[\si^{ik}_{\la\al} (a^{\al\m}_{kj}y^j_\m + b^\al_k)], \label{b4122}
\eea
\bea
&& \Pi=\cR(\Pi)\op\oplus_Y\cP(\Pi)=\Ker\si \op\oplus_Y N_L, \label{N20} \\
&& p^\la_i = \cR^\la_i+\cP^\la_i= [p^\la_i -
a^{\la\m}_{ij}\si^{jk}_{\m\al}p^\al_k] +
[a^{\la\m}_{ij}\si^{jk}_{\m\al}p^\al_k]. \label{N20'}
\eea
\end{theo}

\begin{proof} The proof follows from a direct computation by means of the
relations (\ref{250}), (\ref{N45}) and (\ref{N21}).
\end{proof}

It is readily observed that, with respect to the coordinates $\cS^i_\la$
and $\cF^i_\la$ (\ref{b4122}), the Lagrangian (\ref{N12}) reads 
\beq
\cL=\frac12 a^{\la\m}_{ij}\cF^i_\la\cF^j_\m +c'. \label{cmp31}
\eeq
Note that, in gauge theory, we have the canonical splitting (\ref{N18}) where
$2\cF$ is the strength tensor.$^{3,8,9}$  The
Yang--Mills Lagrangian of gauge theory is exactly of the form (\ref{cmp31})
where $c'=0$. The Lagrangian of Proca fields is also of the form (\ref{cmp31})
where
$c'$ is the mass term. This is an example of a  degenerate
Lagrangian system without gauge symmetries.

Given the linear map $\si$ (\ref{N17}) and a connection $\G$
(\ref{N16}), let us consider the affine Hamiltonian map
\beq
\Phi=\wh\G+\si:\Pi \op\to J^1Y,  \qquad
\Phi^i_\la = \G^i_\la  + \si^{ij}_{\la\m}p^\m_j, \label{N19}
\eeq
and the Hamiltonian form
\ben
&& H=H_\Phi +\Phi^*L= p^\la_idy^i\w\om_\la - [\G^i_\la
(p^\la_i-\frac12 b^\la_i) +\frac12 \si^{ij}_{\la\m}p^\la_ip^\m_j-c]\om=
\label{N22}\\
&& \qquad (\cR^\la_i+\cP^\la_i)dy^i\w\om_\la - [(\cR^\la_i+\cP^\la_i)\G^i_\la
+\frac12
\si^{ij}_{\la\m}\cP^\la_i\cP^\m_j-c']\om.\nonumber
\een

\begin{theo} \label{cmp30}  The Hamiltonian forms (\ref{N22})
spanned by connections $\G$ (\ref{N16}) are associated with the
Lagrangian  (\ref{N12}) and constitute a complete set.
\end{theo}

\begin{proof}
By the very definitions of $\G$ and $\si$, the Hamiltonian map (\ref{N19})
satisfies the condition (\ref{2.30a}). A direct computation shows that
$\Phi=\wh H$. Then the relation (\ref{2.30b}) also holds and, if
$\G$ is a connection (\ref{N16}), the Hamiltonian
form $H$ (\ref{N22}) is associated with the Lagrangian  (\ref{N12}).
Let us write the corresponding Hamilton equations (\ref{b4100a}) for
a section $r$ of the Legendre bundle $\Pi\to X$. They are
\beq
J^1s= (\wh\G+\si)\circ r, \qquad s=\pi_{\Pi Y}\circ r. \label{N29}
\eeq
Due to the surjections $\cS$ and $\cF$ (\ref{N18}),
the Hamilton equations (\ref{N29}) break in two parts
\ben
&&\cS\circ J^1s=\G\circ s, \label{N23}\\
&&\dr_\la r^i-
\si^{ik}_{\la\al} (a^{\al\m}_{kj}\dr_\mu r^j + b^\al_k)=\G^i_\la\circ s,
\nonumber \\
&&\cF \circ J^1s=\si\circ r, \label{N28}\\
&&\si^{ik}_{\la\al} (a^{\al\m}_{kj}\dr_\mu r^j + b^\al_k)=
\si^{ik}_{\la\al}r^\al_k.\nonumber
\een
Let $s$ be an arbitrary section of $Y\to X$,
e.g., a solution of the Euler--Lagrange
equations. There exists a connection $\G$ (\ref{N16}) such
that the relation (\ref{N23}) holds, namely, $\G={\cal S}\circ\G'$ where
$\G'$ is a
connection on $Y\to X$ which has $s$ as an integral section. 
It is easily seen that, in this case, the Hamiltonian map (\ref{N19})
satisfies the relation (\ref{2.36'}) for $s$. 
Hence, the Hamiltonian forms (\ref{N22}) constitute
a complete set. 
\end{proof}

Of course, this complete set is neither minimal nor
unique. Hamiltonian forms $H$ (\ref{N22}) of this set differ from
each other in the term $\f^i_\la\cR^\la_i$ where $\f$ are the soldering forms
(\ref{cmp21}). If follows from the splitting (\ref{N20}) that this term
vanishes on the Lagrangian constraint space.
The corresponding  constrained Hamiltonian form
$H_N=i_N^*H$ and the  constrained Hamilton equations (\ref{N44}) can be
written. In the case of quadratic Lagrangians, we can improve Theorem
\ref{cmp22} as follows.

\begin{theo} \label{cmp23} For every Hamiltonian
form $H$ (\ref{N22}),
the Hamilton equations (\ref{b4100b}) and (\ref{N28}) restricted to the
Lagrangian constraint space $N_L$  are equivalent to the constrained Hamilton
equations.
\end{theo}

\begin{proof} Due to the splitting (\ref{N20}), we have the corresponding 
splitting 
of the vertical tangent bundle $V_Y\Pi$ of the Legendre bundle $\Pi\to Y$.
In particular, any
vertical vector field
$u$ on
$\Pi\to X$ admits the decomposition
\be
&& u= [u-u_{TN}] + u_{TN},  \\
&& u_{TN}=u^i\dr_i +a^{\la\m}_{ij}\si^{jk}_{\m\al}u^\al_k\dr_\la^i, 
\ee
such that $u_N=u_{TN}\mid_{N_L}$ is a vertical vector field on the Lagrangian
constraint space $N_L\to X$. Let us consider the equations
\beq
r^*(u_{TN}\rfloor dH)=0 \label{cmp15}
\eeq
where $r$ is a section of $\Pi\to X$ and $u$ is an arbitrary vertical vector
field on $\Pi\to X$. They are equivalent to the pair of equations
\bea
&& r^*(a^{\la\m}_{ij}\si^{jk}_{\m\al}\dr_\la^i\rfloor dH)=0,
\label{b4125a} \\
&& r^*(\dr_i\rfloor dH)=0. \label{b4125b}
\eea
The equations (\ref{b4125b}) are obviously the Hamilton equations
(\ref{b4100b}) for $H$. Bearing in mind the relations (\ref{250}) and
(\ref{N21}), one can easily show that the equations (\ref{b4125a}) 
coincide with the Hamilton equations (\ref{N28}). The proof is
completed by observing that, restricted to the Lagrangian constraint space
$N_L$, the equations (\ref{cmp15}) are exactly the constrained Hamilton
equations (\ref{N44'}).
\end{proof} 

Note that, in Hamiltonian gauge theory, the restricted Hamiltonian form and
the restricted Hamilton equations are gauge invariant.

Theorem \ref{cmp23} shows that, restricted to the Lagrangian constraint
space, the Hamilton equations for different Hamiltonian forms (\ref{N22})
associated with the same quadratic Lagrangian (\ref{N12}) differ from each
other in the equations (\ref{N23}). These equations are independent of 
momenta and play the role of gauge-type conditions as follows. 

By virtue of Theorem \ref{3.01}, the constrained Hamilton equation are
quasi-equivalent to the Cartan equations. A section
$\ol s$ of
$J^1Y\to X$ is a solution of the Cartan equations for an almost
regular quadratic  Lagrangian (\ref{N12}) iff
$r=\wh L\circ \ol s$ is a solution of the   
Hamilton equations (\ref{b4100b}) and (\ref{N28}). 
In particular, 
let $\ol s$ be such a solution of the Cartan
equations and
$\ol s_0$ a section of the fiber bundle 
$T^*X\op\ot_Y VY\to X$ which takes its values into $\Ker \ol L$ (see
(\ref{N13'})) and projects onto the section 
$s=\pi^1_0\circ \ol s$ of $Y\to X$. Then the affine sum $\ol s +\ol s_0$ over
$s(X)\subset Y$  is also a solution of the Cartan equations.
Thus, we come to the notion of a gauge-type freedom of the Cartan equations for
an almost regular quadratic Lagrangian $L$. One
can speak of the gauge classes of solutions of the Cartan equations whose
elements differ from each other in the above-mentioned sections $\ol s_0$. Let
$z$ be such a gauge class whose elements project onto a section $s$ of $Y\to
X$. For different connections $\G$ (\ref{N16}), we consider the condition
\beq
\cS\circ\ol s=\G\circ s, \qquad \ol s\in z. \label{cmp25}
\eeq

\begin{prop} \label{cmp26} 
(i) If two elements $\ol s$ and $\ol s'$ of the same  gauge class
$z$ obey the same condition (\ref{cmp25}), then $\ol s=\ol s'$.
(ii) For any solution $\ol s$ of the Cartan equations, there exists a
connection (\ref{N16}) which fulfills the condition (\ref{cmp25}).
\end{prop}
 
\begin{proof}
(i) Let us consider the affine difference  $\ol s-\ol s'$ over
$s(X)\subset Y$. We have $\cS(\ol s-\ol s')=0$ iff $\ol s=\ol s'$.
(ii) In the proof of Theorem \ref{cmp30}, we have shown that, given
$s=\pi^0_1\circ\ol s$, there exists a connection
$\G$ (\ref{N16}) which fulfills the relation (\ref{N23}). Let us consider the
affine difference $\cS(\ol s- J^1s)$ over $s(X)\subset Y$. This is a local
section of the vector bundle $\Ker \ol L\to Y$ over $s(X)$. Let $\f$ be its 
prolongation onto $Y$. It is easy to see that $\G+\f$ is the
desired connection.
\end{proof}

Due to the properties in Proposition \ref{cmp26}, one can treat 
(\ref{cmp25}) as a gauge-type condition on solutions
of the Cartan equations. The Hamilton equations (\ref{N23}) exemplify this
gauge-type condition when
$\ol s= J^1s$ is a solution of the Euler--Lagrange equations. At the same
time, the above-mentioned freedom characterizes solutions of the Cartan
equations, but not the Euler--Lagrange ones. First of all, this freedom
reflects the degeneracy of the Cartan equations (\ref{b336a}). Therefore, in
the Hamiltonian gauge theory, the above freedom is not related directly to the
familiar gauge invariance. Nevertheless, the Hamilton equations (\ref{N23})
are not gauge invariant, and also can play the role of gauge conditions in
gauge theory. Indeed, given a Hamiltonian form $H$ (\ref{N22}), the
corresponding Lagrangian $L_H$ (\ref{Q3}) reads
\beq
\cL_H=\cR^\la_i(\cS^i_\la-\G^i_\la) +\cP^\la_i\cF^\la_i
-\frac12\si^{ij}_{\la\m}\cP^\la_i\cP^\m_j +c'. \label{cmp86}
\eeq
In comparison with the Lagrangian $L$ (\ref{N12}) and the constrained
Lagrangian $L_H\mid_{J^1N_L}$,
the Lagrangian (\ref{cmp86}) includes the additional gauge fixing term
$\cR^\la_i(\cS^i_\la-\G^i_\la)$.

\bigskip 

\noindent 
{\bf VII. VERTICAL EXTENSION OF POLYSYMPLECTIC FORMALISM}
\bigskip

The extension of polysymplectic 
formalism to the vertical tangent bundle $VY$ of $Y\to X$ is a preliminary
step toward its BRST extension. 
The Legendre bundle (\ref{00}) over $VY\to X$, called the vertical Legendre
bundle, is
\be
\Pi_{VY}=V^*VY\op\w_{VY}(\op\w^{n-1} T^*X). 
\ee
 We will use the compact notation
\be
\dot\dr_i=\frac{\dr}{\dr\dot y^i}, \qquad \dot\dr^i_\la=\frac{\dr}{\dr\dot
p_i^\la}, \qquad \dr_V=\dot y^i\dr_i + 
\dot p^\la_i \dr_\la^i.
\ee

\begin{lem}
There exists the bundle isomorphism 
\beq
\Pi_{VY}\op\cong_{VY} V\Pi,
\qquad p^\la_i\llra\dot p^\la_i, \qquad q^\la_i\llra p^\la_i, \label{cmp42}
\eeq
written relative to the holonomic coordinates $(x^\la, y^i, \dot y^i,
p^\la_i, q^\la_i)$ on $\Pi_{VY}$ and  $(x^\la, y^i, p^\la_i,
\dot y^i,\dot p^\la_i)$ on $V\Pi$.
\end{lem}

\begin{proof}
Similar to the well-known isomorphism between the fiber bundles $TT^*X$ and
$T^*TX$,$^5$  the isomorphism 
\be
VV^*Y \op\cong_{VY} V^*VY, \quad p_i\llra\dot v_i, 
\quad \dot p_i\llra\dot y_i, 
\ee
can be established by inspection of the
transformation laws of the holonomic
coordinates $(x^\la, y^i, p_i)$ on
$V^*Y$ and $(x^\la, y^i, v^i)$ on $VY$.
\end{proof} 

It follows that 
Hamiltonian formalism on the vertical Legendre bundle $\Pi_{VY}$ can be
developed as the vertical extension onto
$V\Pi$ of Hamiltonian formalism on $\Pi$, where
the canonical conjugate pairs are
$(y^i,\dot p^\la_i)$ and $(\dot y^i,p_i^\la)$. In particular, 
due to the isomorphism (\ref{cmp42}),
$V\Pi$ is endowed with the canonical polysymplectic form 
(\ref{406}) which reads
\beq
\Om_{VY}=[d\dot p^\la_i\w dy^i +dp^\la_i\w d\dot y^i]\w\om\ot\dr_\la.
\label{cmp35}
\eeq

Let $Z_{VY}$ be the homogeneous Legendre bundle (\ref{N41}) over $VY$ 
with the corresponding coordinates $(x^\la,y^i,\dot
y^i,p_i^\la,q_i^\la,p)$. It can be endowed with  
the canonical form $\Xi_{VY}$ (\ref{N43}). Sections of the
affine bundle
\beq
 Z_{VY}\to V\Pi, \label{cmp41}
\eeq
by definition, provide Hamiltonian forms on $V\Pi$. Let us
consider the following particular case of these forms which are
related to those on the Legendre bundle $\Pi$.
Due to the fiber bundle
\ben
&&\zeta: VZ_Y\to Z_{VY}, \label{cmp40}\\
&& (x^\la,y^i,\dot y^i,p_i^\la,q_i^\la,p) \circ\zeta= (x^\la,y^i,\dot
y^i,\dot p_i^\la, p_i^\la,\dot p), \nonumber
\een
the vertical tangent bundle $VZ_Y$ of $Z_Y\to X$ is provided with
the exterior form 
\be
\Xi_V=\zeta^*\Xi_{VY}= \dot p\om + (\dot p^\la_i
dy^i + p^\la_id\dot y^i) \w\om_\la.
\ee
Given the affine bundle 
$Z_Y\to\Pi$ (\ref{b418'}), we have the fiber bundle
\beq
V\pi_{Z\Pi}: VZ_Y\to V\Pi, \label{cmp34}
\eeq
where $V\pi_{Z\Pi}$ is the vertical tangent map to $\pi_{Z\Pi}$. The fiber
bundles (\ref{cmp41}), (\ref{cmp40}) and (\ref{cmp34}) form the commutative
diagram.

Let $h$ be a section of the affine bundle $Z_Y\to \Pi$ and
$H=h^*\Xi$ the corresponding Hamiltonian form (\ref{b418}) on $\Pi$.  
Then the section $Vh$ of the fiber bundle (\ref{cmp34}) and the
corresponding section $\zeta\circ Vh$ of the affine bundle (\ref{cmp41})
defines the Hamiltonian form
\ben
&& H_V=(Vh)^*\Xi_V =(\dot p^\la_idy^i + p^\la_i d\dot
y^i)\w\om_\la -\cH_V\om,
\label{m17}\\ 
&& \cH_V =\dr_V\cH=(\dot y^i\dr_i +\dot
p^\la_i\dr^i_\la)\cH, \nonumber
\een
on $V\Pi$. It is called the vertical extension of $H$.
In particular, given the splitting (\ref{4.7}) of $H$ with respect to
a connection
$\G$ on $Y\to X$, we have the corresponding splitting
\be
\cH_V=\dot p^\la_i\G^i_\la +\dot y^j p^\la_i\dr_j\G^i_\la +\dr_V\wt\cH_\G
\ee
of $H_V$ with respect to the vertical connection $V\G$ (\ref{cmp43}) on $VY\to
X$.

\begin{prop} 
Let $\g$ (\ref{cmp33}) be a Hamiltonian connection on $\Pi$
associated with a Hamiltonian form $H$.
Then its vertical prolongation $V\g$ 
(\ref{cmp43}) on $V\Pi\to X$ is a Hamiltonian connection associated
with the vertical Hamiltonian form $H_V$ (\ref{m17}).
\end{prop}

\begin{proof}
The proof follows from a direct computation. We have
\be
V\g=\g + dx^\m\ot [\dr_V\g^i_\m\dot\dr_i +\dr_V\g^\la_{\m i}\dot\dr_\la^i].
\ee
Components of this connection obey the Hamilton equations (\ref{3.10}) 
and the equations
\beq
\dot \g^i_\m=\dr^i_\m\cH_V=\dr_V\dr^i_\m\cH,\qquad
 \dot \g^\la_{\la i}=-\dr_i\cH_V=-\dr_V\dr_i\cH. \label{cmp51}
\eeq
\end{proof}

 In order to clarify the physical meaning of
the Hamilton equations (\ref{cmp51}), let us suppose that
$Y\to X$ is a vector bundle.  Given a solution $r$ of the
Hamilton equations for $H$, let 
$\ol r$ be a Jacobi field,  i.e., $r+\ve \ol r$
is also a solution of the same Hamilton equations modulo terms of order $>1$ in
$\ve$. Then it is readily observed that the Jacobi field $\ol r$ satisfies
the Hamilton equations (\ref{cmp51}). At the same time, the
Lagrangian $L_{H_V}$ (\ref{Q3}) on $J^1V\Pi$, defined by the Hamiltonian form
$H_V$ (\ref{m17}),  takes the form
\beq
\cL_{VH}=h_0(H_V)=\dot p^\la_i(y^i_\la- \dr^i_\la\cH) -\dot y^i(p^\la_{\la
i} + \dr_i\cH) +d_\la(p^\la_i\dot y^i), \label{cmp105}
\eeq
where $\dot p^\la_i$, $\dot y^i$ play the role of Lagrange multipliers. The
corresponding generating functional reduces to Dirac's
$\delta$-functions at classical solutions.
\bigskip 

\noindent 
{\bf VIII. BRST-EXTENDED HAMILTONIAN FORMALISL}
\bigskip

The BRST extension of Hamiltonian mechanics$^{11,12}$  shows
that:  (i) one should consider vector bundles $Y\to X$ in order to introduce
generators of BRST and anti-BRST transformations, and (ii) one can narrow the
class of superfunctions under consideration because the BRST extension
of a Hamiltonian is a polynomial of a finite degree in odd variables.
Therefore, we will formulate the BRST extension on the polysymplectic
Hamiltonian formalism in the terms of simple graded manifolds.

Recall$^{24,25}$ that by a graded manifold is meant the pair $(Z,\cA)$ of a
smooth manifold $Z$ and a sheaf $\cA$ of graded-commutative $\bR$-algebras such
that

(i) there is the exact sequence of sheaves
\beq
0\to \cJ \to\cA \to C^\infty(Z)\to 0, \qquad
\cJ=\cA_1+(\cA_1)^2,\label{cmp140}
\eeq

(ii) $\cJ/\cJ^2$ is a locally free
$C^\infty(Z)$-module of finite rank, and $\cA$ is locally isomorphic to the
exterior bundle $\op\w_{C^\infty(Z)}(\cJ/\cJ^2)$. The exact sequence
(\ref{cmp140}) admits the canonical splitting $C^\infty(Z)\to\cA$, and the
well-known Batchelor's theorem takes place.

\begin{theo}$^{25,26}$
Let $(Z,\cA)$ be a graded manifold. There exists a vector bundle 
$E\to Z$ with an $m$-dimensional
typical fiber $V$ such that $\cA$ is isomorphic
to the sheaf $\cA_E$ of sections of the exterior bundle
\beq
\w E^*=\bR\op\oplus_Z(\op\oplus_{k=1}^m\op\w^k E^*) \label{z780}
\eeq
whose typical fiber is the finite Grassman algebra $\w V^*$.
\end{theo}

This isomorphism fails to be canonical, and restricts 
transformations of a graded manifold to those induced by the bundle
automorphisms of $E\to Z$. Nevertheless, this class of transformations is
sufficient for our purposes because we consider the graded extension of
Hamiltonian formalism on smooth manifolds when the vector bundle $E$
(\ref{cmp68}) below is fixed. We will call $(Z,\cA_E)$  the simple graded
manifold. This is not the terminology of Ref. \cite{cari97} where this term
is applied to all finite graded manifolds, but in connection with 
Batchelor's isomorphism.

Global sections of the exterior bundle (\ref{z780}) are called
superfunctions due to the equivalence between the graded manifolds $(Z,\cA)$
and the De Witt supermanifolds whose body is $Z$.$^{25,28}$ This isomorphism
is important for for functional integration over superfunctions. Superfunctions
make up a
$\bZ_2$-graded ring $\cA_E(X)$.  
Let $\{c^a\}$ be the holonomic bases for $E^*\to Z$ with respect to some bundle
atlas with transition functions $\{\rho^a_b\}$, i.e.,
$c'^a=\rho^a_b(z)c^b$. Then superfunctions read
\beq
f=\op\sum_{k=0}^m \frac1{k!}f_{a_1\ldots
a_k}c^{a_1}\cdots c^{a_k}, \label{z785}
\eeq
where $f_{a_1\cdots
a_k}$ are local functions on $Z$, and we omit the symbol of exterior product of
elements $c$. The coordinate transformation law of superfunctions
(\ref{z785}) is obvious. We will use the
notation $\nw .$ of the Grassman parity.

Given a graded manifold $(Z,\cA)$, the sheaf $\der\cA$ of graded derivations
of $\cA$ is introduced. This is a subsheaf of endomorphisms of $\cA$ whose
sections
$u$ on an open subset $U\subset Z$ are graded derivations of the restriction 
$\cA\mid_U$ of the sheaf $\cA$ to $U$, i.e.,  
\be
 u(ff')=u(f)f'+(-1)^{\nw u\nw f}fu (f') 
\ee
for the homogeneous elements $u\in(\der\cA)(U)$ and $f,f'\in \cA\mid_U$.
In the case of graded manifolds, derivations of $\cA$ are local operators. It
means that $(\der\cA)(U)=\der\cA(U)$, i.e., if $U'\subset U$ are open
sets, there is the restriction morphism $\der\cA(U)\to
\der\cA(U')$.$^{25}$ It follows that the sheaf $\der\cA$ coincides with the
sheaf of graded $\cA$ modules $U\to\der\cA(U)$.
Its sections are 
called supervector fields on a manifold $Z$.
The dual of the sheaf $\der\cA$ is the sheaf
 $\der^*\cA$ generated by the $\cA$-linear morphisms
\beq
\f:\der\cA(U)\to \cA_U. \label{z789}
\eeq
One can think of its sections as being 1-superforms on a manifold $Z$. 

In the case of a simple graded manifold $(Z,\cA_E)$ supervector fields and
1-superforms can be represented by sections of vector bundles as follows.
Due to the canonical splitting
$VE= E\times E$, the vertical tangent bundle 
$VE\to E$ can be provided with the fiber bases $\{\dr_a\}$ dual of $\{c^a\}$.
These are fiber bases for $\pr_2VE=E$. Let $(z^A)$ be coordinates on $Z$. Then
a supervector field on a trivialization domain $U$ read
$u= u^A\dr_A + u^a\dr_a$
where $u^A, u^a$ are local superfunctions.
It yields a graded endomorphism of $\cA_E(U)$ by the rule
\beq
u(f_{a\ldots b}c^a\cdots c^b)=u^A\dr_A(f_{a\ldots b})c^a\cdots c^b +u^a
f_{a\ldots b}\dr_a\rfloor (c^a\cdots c^b). \label{cmp50'}
\eeq
This implies the corresponding
coordinate transformation law 
\be
u'^A =u^A, \qquad u'^a=\rho^a_ju^j +u^A\dr_A(\rho^a_j)c^j
\ee
of supervector fields. It follows that supervector fields on $Z$, which we
agree to call $E$-determined supervector fields, can be represented by
sections of the vector bundle
$\cV_E\to Z$ which is locally isomorphic to the vector bundle
\be
\cV_E\mid_U\approx\w E^*\op\ot_Z(\pr_2VE\op\oplus_Z TZ)\mid_U,
\ee
and has the transition functions
\be
&& z'^A_{i_1\ldots i_k}=\rho^{-1}{}_{i_1}^{a_1}\cdots
\rho^{-1}{}_{i_k}^{a_k} z^A_{a_1\ldots a_k}, \\
&& v'^i_{j_1\ldots j_k}=\rho^{-1}{}_{j_1}^{b_1}\cdots
\rho^{-1}{}_{j_k}^{b_k}\left[\rho^i_jv^j_{b_1\ldots b_k}+ \frac{k!}{(k-1)!} 
z^A_{b_1\ldots b_{k-1}}\dr_A(\rho^i_{b_k})\right] 
\ee
of the bundle coordinates $(z^A_{a_1\ldots a_k},v^i_{b_1\ldots b_k})$,
$k=0,\ldots,m$. These transition functions
fulfill the cocycle relations. 
There is the exact sequence over $Z$ of vector
bundles
\beq
0\to \w E^*\op\ot_Z\pr_2VE\to\cV_E\to \w E^*\op\ot_Z TZ\to 0. \label{cmp92}
\eeq
Due to the above mentioned locality property the 
sheaf of sections of the vector bundle
$\cV_E\to Z$ is isomorphic to the sheaf $\der\,\cA_E$. Global sections of 
$\cV_E\to Z$ constitute the $\cA_E(Z)$-module of supervector fields on $Z$,
which is also a Lie
superalgebra with respect to the bracket 
\be
[u,u']=uu' + (-1)^{\nw u\nw{u'}+1}u'u.
\ee

One can think of a splitting 
\beq
\wt\g:\dot z^A\dr_A \mapsto \dot z^A(\dr_A +\wt\g_A^a\dr_a) \label{cmp70}
\eeq
of the exact sequence (\ref{cmp92}) as being a graded connection, though
this is not a true connection on $\cV_E\to Z$. A
graded connection can be represented by a section
\beq
\wt \g=dz^A\ot(\dr_A +\wt\g^a_A\dr_a) \label{cmp93}
\eeq
of the vector bundle $T^*Z\op\ot_Z\cV_E\to Z$ such that the composition
\be
Z\op\to^{\wt\g}T^*Z\op\ot_Z\cV_E\to T^*Z\op\ot_Z (\w
E^*\op\ot_Z TZ)\to T^*Z\op\ot_ZTZ
\ee
is the canonical form $dz^A\ot\dr_A$ on $Z$.
Such a graded connection $\wt\g$ transforms every vector field $\tau$ on $Z$
into a supervector field 
\be
\tau=\tau^A\dr_a\mapsto \wt\g\tau=\tau^A(\dr_A +\wt\g_A^a\dr_a),
\ee
and provides the corresponding decomposition
\be  
u= u^A\dr_A + u^a\dr_a=u^A(\dr_A +\wt\g_A^a\dr_a) + (u^a-
u^A\wt\g_A^a)\dr_a
\ee
of supervector fields on $Z$. 
For instance, every linear connection 
\be
\g=dz^A\ot (\dr_A +\g_A{}^a{}_bv^b\dr_a) 
\ee
on the vector bundle $E\to Z$ defines the graded connection 
\beq
\g_S=dz^A\ot (\dr_A +\g_A{}^a{}_bc^b\dr_a) \label{cmp73}
\eeq
such that, for any vector field $\tau$ on $Z$ and any superfunction $f$,
the graded derivation $\g_S\tau(f)$ is exactly the covariant derivative
$\tau^A\nabla_A f$ relative to the connection $\g$.

\begin{rem}
Let now $Z\to X$ be a fiber bundle, coordinated by $(x^\la,z^i)$. 
Let 
\be
\g=\G +\g_\la{}^a{}_bv^bdx^\la\ot\dr_a
\ee
be a connection on $E\to X$ which is a linear morphism over a connection $\G$
on $Z\to X$. Then we have the bundle monomorphism
\be
\g_S: \w E^*\op\ot_ZTX\ni u^\la\dr_\la \mapsto u^\la(\dr_\la+\G^i_\la\dr_i
+\g_\la{}^a{}_bc^b\dr_a)\in \cV_E 
\ee
over $Z$, called a composite graded connection on $Z\to X$.
It is represented by a section
\beq
\g_S= \G + \g_\la{}^a{}_bc^bdx^\la\ot\dr_a \label{cmp122}
\eeq
of the fiber bundle $T^*X\op\ot_Z\cV_E\to Z$ such that the composition
\be
Z\op\to^{\g_S}T^*X\op\ot_Z\cV_E\to T^*X\op\ot_Z (\w
E^*\op\ot_Z TZ)\to T^*X\op\ot_ZTX
\ee
is the pull-back onto $Z$ of the canonical form $dx^\la\ot\dr_\la$ on $X$.
\end{rem}

The $\w E^*$-dual $\cV^*_E$ of $\cV_E$ is a vector bundle over $Z$
which is locally isomorphic to the vector bundle
\be
\cV^*_E\mid_U\approx \w E^*\op\ot_Z(\pr_2VE^*\op\oplus_Z T^*Z)\mid_U,
\ee
and has the transition functions
\be
&& v'_{j_1\ldots j_kj}= \rho^{-1}{}_{j_1}^{a_1}\cdots
\rho^{-1}{}_{j_k}^{a_k} \rho^{-1}{}_j^a v_{a_1\ldots a_ka}, \nonumber\\
&& z'_{i_1\ldots i_kA}=
\rho^{-1}{}_{i_1}^{b_1}\cdots
\rho^{-1}{}_{i_k}^{b_k}\left[z_{b_1\ldots b_kA}+ \frac{k!}{(k-1)!} 
v_{b_1\ldots b_kj}\dr_A(\rho^j_{b_k})\right] 
\ee
of the bundle coordinates $(z_{a_1\ldots a_kA},v_{b_1\ldots b_kj})$,
$k=0,\ldots,m$, with respect to the dual bases $\{dz^A\}$ for $T^*Z$ and
$\{dc^b\}$ for $\pr_2V^*E=E^*$. There is the exact sequence
\beq
0\to \w E^*\op\ot_ZT^*Z\to\cV^*_E\to \w E^*\op\ot_Z \pr_2VE^*\to 0.
\label{cmp72}
\eeq
The sheaf of sections of $\cV^*_E\to Z$ is isomorphic to
the sheaf $\der^*\cA_E$. Global sections of the vector bundle
$\cV^*\to Z$ constitute the $\cA_E(Z)$-module of $E$-determined exterior
1-superforms $\f=\f_A dz^A + \f_adc^a$
on $Z$ with the coordinate transformation law
\be
\f'_a=\rho^{-1}{}_a^b\f_b, \qquad \f'_A=\f_A
+\rho^{-1}{}_a^b\dr_A(\rho^a_j)\f_bc^j.
\ee
Then
the morphism (\ref{z789}) can be seen as the interior product 
\beq
u\rfloor \f=u^A\f_A + (-1)^{\nw{\f_a}}u^a\f_a. \label{cmp65}
\eeq
Any graded connection $\wt\g$ (\ref{cmp93}) also yields the
splitting of the exact sequence (\ref{cmp72}), and defines the corresponding
decomposition of 1-superforms
\be
\f=\f_A dz^A + \f_adc^a =(\f_A+\f_a\wt\g_A^a)dz^A +\f_a(dc^a
-\wt\g_A^adz^A). 
\ee

Accordingly, $k$-superforms $\f$ are sections
of the graded exterior bundle $\ol\w^k_Z\cV^*_E$ such that
\be
 \f\ol\w\si =(-1)^{\nm\f\nm\si +\nw\f\nw\si}\si\ol\w \f.  
\ee
The interior product (\ref{cmp65})
is extended to higher degree superforms by the rule  
\be
u\rfloor (\f\ol\w\si)=(u\rfloor \f)\ol\w \si
+(-1)^{\nm\f+\nw\f\nw{u}}\f\ol\w(u\rfloor\si). 
\ee
Recall that the graded exterior differential
$d$ of superfunctions is introduced in accordance with the condition 
$u\rfloor df=u(f)$
for an arbitrary supervector field $u$, and  is
extended uniquely to higher degree superforms by the rules
\be
d(\f\ol\w\si)= (d\f)\ol\w\si +(-1)^{\nm\f}\f\ol\w(d\si), \qquad  d\circ d=0.
\ee
It takes the coordinate form
\be
d\f= dz^A \ol\w \dr_A(\f) +dc^a\ol\w \dr_a(\f), 
\ee
where the left derivatives 
$\dr_A$, $\dr_a$ act on the coefficients of superforms by the rule
(\ref{cmp50'}), and they are graded commutative with the forms $dz^A$, $dc^a$.
The Lie
derivative of a superform $\f$ along a supervector field $u$ is given by
the familiar formula
\beq
\bL_u\f= u\rfloor d\f + d(u\rfloor\f). \label{cmp66}
\eeq

\begin{rem} \label{cmp102}
Given a vector bundle $E\to Z$, let us consider the jet manifold $J^1E$,
coordinated by $(z^A,v^a,v^a_A)$. This is also a vector bundle over $Z$. Then
one can construct the corresponding fiber bundles $\cV_{J^1E}$ and
$\cV_{J^1E}^*$. Due to the monomorphism $E^*\to (J^1E)^*$, there is
the monomorphism
$\cV_E^*\to
\cV_{J^1E}^*$, i.e., every
$E$-determined superform on $Z$ can be also seen as a $J^1E$-determined
superform. In particular, the horizontal projection
$h_0$ (\ref{cmp100}) gives rise to the 0-graded homomorphism
\beq
h_0: dc^a\to c^a_Adz^A \label{cmp101}
\eeq
which sends $E$-determined superforms onto horizontal $J^1E$-determined
superforms.
\end{rem}

Turn now to the BRST extension of covariant Hamiltonian formalism on the
Legendre bundle $\Pi$ (\ref{00}) when $Y\to X$ is a vector bundle. Let us
apply the above construction of simple graded manifolds to the case of the
vertical tangent bundle 
\beq
E=VV\Pi=V\Pi\op\oplus_X V\Pi\ar^{\pr_1} V\Pi \label{cmp68}
\eeq
over the vertical Legendre bundle $Z=V\Pi\to X$. 
Let $(x^\la, y^i, p^\la_i,\dot y^i,\dot p^\la_i)$ be the holonomic
coordinates on $V\Pi$. Then the dual $E^*$ of $E$ can be endowed with the
associated fiber bases
$\{c^i,c_i^\la,\ol c^i,\ol c_i^\la\}$ such that
$c^i$ and $\ol c^i$ have the same linear coordinate
transformation law as the coordinates $y^i$ and $\dot y^i$, while
$c_i^\la$  and $\ol c_i^\la$ have
those of the coordinates $p_i^\la$ and $\dot p_i^\la$. The corresponding
supervector fields and superforms are introduced on $V\Pi$ as sections of the
vector bundles $\cV_{VV\Pi}$ and $\cV^*_{VV\Pi}$, respectively. Let us
complexify these bundles as $\bC\op\ot_X\cV_{VV\Pi}$ and
$\bC\op\ot_X\cV^*_{VV\Pi}$.  

As in mechanics, the main criterion of the BRST extension of covariant
Hamiltonian formalism is its invariance under  BRST and anti-BRST
transformations whose generators are the supervector fields
\ben
&& \vt_Q= \dr_c +i\dot y^i
\frac{\dr}{\dr \ol c^i} + i\dot p_i^\la\frac{\dr}{\dr\ol c_i^\la},
\quad  \vt_{\ol Q}=\dr_{\ol c} -i\dot y^i
\frac{\dr}{\dr c^i} - i\dot p_i^\la\frac{\dr}{\dr c_i^\la}, \label{z746}\\
&& \dr_c=c^i\dr_i + c_i^\la\dr^i_\la, \qquad
  \dr_{\ol c}=\ol c^i\dr_i + \ol c_i^\la\dr^i_\la, \nonumber
\een
on $V\Pi$. They fulfill the nilpotency rules
\be
\vt_Q\vt_Q=0, \qquad \vt_{\ol Q} \vt_{\ol Q}=0, \qquad \vt_{\ol Q} \vt_Q
+\vt_Q\vt_{\ol Q}=0.
\ee

The BRST- and
anti-BRST-invariant extension of the polysymplectic form  
$\Om_{VY}$ (\ref{cmp35}) on $V\Pi$ is the
$TX$-valued superform
\be
\Om_S=[d\dot p_i^\la\w dy^i +dp_i^\la\w d\dot y^i +i(d\ol c_i^\la\w dc^i- d\ol
c^i\w dc_i^\la)]\w \om\ot\dr_\la
\ee
on $V\Pi$, where $(c^i,-i\ol c^\la_i)$ and $(\ol c^i,i c^\la_i)$ are the
conjugate pairs.
Let $\g$ be a Hamiltonian connection for a Hamiltonian form $H$. The double
vertical connection $VV\g$ on $VV\Pi\to X$ is a linear morphism over the
vertical connection $V\g$ on $V\Pi\to X$, and so defines the composite
graded connection 
\be
(VV\g)_S =V\g + dx^\m\ot[\ol g^i_\m\frac{\dr}{\dr \ol c^i} +\ol
g^\la_{\m i}\frac{\dr}{\dr\ol c_i^\la} + g^i_\m\frac{\dr}{\dr c^i} +g^\la_{\m
i}\frac{\dr}{\dr c_i^\la}]
\ee
(\ref{cmp122}) on $V\Pi\to X$, whose components  $g$ and $\ol g$ are given by
the expressions
\be
\ol g^i_\la=\dr_{\ol c}\dr^i_\la\cH, \quad
\ol g^\la_{\la i}=-\dr_{\ol c}\dr_i\cH,\quad
g^i_\la= \dr_c\dr^i_\la\cH,
\quad g^\la_{\la i}= -\dr_c\dr_i\cH.
\ee
This composite graded connection satisfies the relation
\be
(VV\g)_S\rfloor\Om_S=-dH_S,
\ee
and so is a Hamiltonian graded connection for the Hamiltonian
superform 
\beq
H_S=[\dot p_i^\la dy^i +p_i^\la d\dot y^i +i(\ol c_i^\la dc^i +d\ol
c^i c_i^\la)]\om_\la -(\dr_V 
+ i\dr_{\ol c}\dr_c)\cH\om \label{z750}
\eeq
on $V\Pi$. This superform is BRST- and anti-BRST-invariant, i.e., $\bL_\vt
H_S=0$. Thus, it is the desired BRST extension of the Hamiltonian form $H$.

The Hamiltonian superform $H_S$ (\ref{z750}) defines the
corresponding BRST-extension of the Lagrangian $L_H$ (\ref{Q3}). Following
Remark \ref{cmp102}, let us consider the vector bundle
\be
J^1(VV\Pi)=VJ^1(V\Pi)\to J^1(V\Pi)=VJ^1\Pi
\ee
and the corresponding fiber bundle $\cV_{VJ^1(V\Pi)}\to VJ^1\Pi$. It is
readily observed that $VV\Pi$-determined superforms on $V\Pi$ can be seen as
particular $VJ^1(V\Pi)$-determined superforms on $VJ^1\Pi$. Moreover,
combining the horizontal projections $h_0$ (\ref{cmp100}) and (\ref{cmp101})
for exterior forms and superforms, we obtain the 0-graded homomorphism
$h_0$ which sends $VV\Pi$-determined superforms on $V\Pi$ onto the horizontal
$VJ^1(V\Pi)$-determined superforms on $VJ^1\Pi\to X$. Then the horizontal 
superdensity
\ben
&& L_{SH}=h_0(H_S)= L_{VH} + i[(\ol c_i^\la
c^i_\la+\ol c^i_\la c_i^\la)  -\dr_{\ol c}\dr_c\cH]\om=  L_{VH}
+\label{cmp104}\\ 
&& \qquad i[\ol c^\la_i(c^i_\la -\dr_c\dr^i_\la\cH) + (\ol
c^i_\la -\dr_{\ol c}\dr^i_\la\cH)c^\la_i + \ol
c^\la_ic^\m_j\dr^i_\la\dr^j_\m\cH -
\ol c^ic^j\dr_i\dr_j\cH]\om
\nonumber
\een
on $VJ^1\Pi\to X$ can be treated as the desired BRST extension of the
Lagrangian $L_H$ (\ref{Q3}). Note that, in
comparison with the Lagrangian $L_{VH}$ (\ref{cmp105}), the generating
functional determined by the BRST-extended Lagrangian
(\ref{cmp104}) is not reduced to $\delta$-functions.

The BRST-extended Lagrangian $L_{SH}$ (\ref{cmp104}) is also invariant under
the jet prolongations 
\be
J^1\vt=\vt^a\dr_a+ d_\la\vt^a\dr_a^\la
\ee
of the BRST and anti-BRST transformations (\ref{z746}).
Moreover, it is easily verified that both
$L_{HS}$ and
$H_S$ are invariant under transformations whose generators are the
supervector fields
\ben
&& \vt_K=c_i^\la\frac{\dr}{\dr \ol c_i^\la} +c^i\frac{\dr}{\dr \ol c^i}, \quad
\vt_{\ol K}=\ol c_i^\la\frac{\dr}{\dr c_i^\la} +\ol c^i\frac{\dr}{\dr c^i},
\label{cmp130}\\
&& \vt_C=c_i^\la\frac{\dr}{\dr c_i^\la} +c^i\frac{\dr}{\dr c^i} -
\ol c_i^\la\frac{\dr}{\dr \ol c_i^\la} -\ol c^i\frac{\dr}{\dr\ol c^i}.
\nonumber
\een
The supervector fields (\ref{z746}) and (\ref{cmp130}) constitute the 
Lie superalgebra of the well-known group ISp(2):
\ben
&& [Q,Q]=[\ol Q,\ol Q]=[\ol Q, Q]=[K,Q]=[\ol K,\ol Q]=0, \label{cmp128} \\
&& [K,\ol Q]=Q, \quad [\ol K, Q]=\ol Q, \quad [K,\ol K]= C, \quad
[C,K]=2K, \quad [C,\ol K]=-2\ol K. \nonumber
\een

Similarly to the lift $\wt u$ (\ref{cmp90}) onto $\Pi$ of a vector field $u$
on $Y$,  the supervector fields $\vt$ (\ref{z746}) and (\ref{cmp130}) can be
represented as the corresponding graded lift
\be
\vt=\wt u=u^a\dr_a - (-1)^{[y^a]([p_b] +[u^b])}\dr_au^b\frac{\dr}{\dr p_a}
\ee
of some $VVY$-determined supervector fields $u$ on $VY$ which are sections
of the fiber bundle $\cV_{VVY\to VY}$. These supervector fields $u$ read
\ben
&& u_Q= c^i\dr_i + i\dot y^i
\frac{\dr}{\dr \ol c^i},
\quad  u_{\ol Q}=\ol c^i\dr_i  -i\dot y^i
\frac{\dr}{\dr c^i}, \label{cmp131} \\
&& u_K=c_i^\la\frac{\dr}{\dr \ol c_i^\la}, \quad
u_{\ol K}=\ol c_i^\la\frac{\dr}{\dr c_i^\la},
\quad u_C=c^i\frac{\dr}{\dr c^i}  -\ol c^i\frac{\dr}{\dr\ol c^i}.
\nonumber
\een
They also constitute the Lie superalgebra (\ref{cmp128}).
Then by analogy with (\ref{cmp129}), we obtain the corresponding
supercurrents
$J_\vt=\vt\rfloor H_S=u\rfloor H_S$. These are the horizontal
$(n-1)$-superforms
\be
&& Q=(c^i\dot p^\la_i-\dot y^i c_i^\la)\om_\la, \quad \ol Q=(\ol c^i\dot
p^\la_i-\dot y^i \ol c_i^\la)\om_\la, \\
&& K=-i c_i^\la c^i\om_\la, \quad \ol K=i\ol c_i^\la \ol c^i\om_\la,\quad
C=i(\ol c^\la c^i -\ol c^i c_i)\om_\la
\ee
on $V\Pi$. They form the Lie superalgebra (\ref{cmp128}) with respect to the
product (\ref{c2}). It should be emphasized that 
the Lie superalgebra (\ref{cmp128}) provides the canonical symmetries of
any BRST-extended Hamiltonian system.

The following construction is similar to that is met in supersymmetric
mechanics and BRST mechanics.
Given a function $F$ on the Legendre bundle $\Pi$,  
let us consider the operators
\beq
F_\bt=e^{\bt F}\circ\vt\circ e^{-\bt F} =\vt
-\bt\dr_c F, \quad 
\ol F_\bt=e^{-\bt F}\circ\ol\vt\circ e^{\bt F}=\ol\vt
+\bt\dr_{\ol c} F, \quad \bt>0, \label{cmp135}
\eeq
called the  BRST and anti-BRST charges,  which act on superfunctions
on $V\Pi$. These operators are nilpotent, i.e.,
\beq
F_\bt\circ F_\bt=0, \qquad \ol F_\bt\circ \ol F_\bt=0. \label{z763}
\eeq
By the BRST- and anti-BRST-invariant extension of $F$ is meant the
superfunction
\beq
 F_S= -\frac{i}{\bt}(\ol F_\bt\circ F_\bt +F_\bt\circ \ol F_\bt).
\label{cmp133}
\eeq
We have the relations 
\be
F_\bt\circ F_S- F_S\circ F_\bt=0, \qquad
\ol F_\bt\circ F_S- F_S\circ\ol F_\bt=0. 
\ee
These relations together with the relations (\ref{z763}) provide the
operators
 $F_\bt$, $\ol F_\bt$, and $F_S$ with the structure of the Lie
superalgebra sl(1/1).$^{29}$

Let now $\G^i_\la=\G_\la{}^i{}_jy^j$ be a linear
connection  on $Y\to X$ and $\wt\G$ some Hamiltonian connection on
$\Pi\to X$ for the Hamiltonian form $H_\G$ (\ref{3.6}).
Given the  splitting
(\ref{4.7}) of the Hamiltonian form
$H$ with respect to the connection $\G$, there is the corresponding splitting
of the BRST-extended Hamiltonian form
\be
\cH_S=\cH_{\G S} +\wt\cH_{S\G}=\dot p_i^\la\G_\la{}^i{}_jy^j +\dot
y^jp_i^\la\G_\la{}^i{}_j +i(\ol c_i^\la\G_\la{}^i{}_jc^j +\ol
c^j\G_\la{}^i{}_jc_i^\la) +(\dr_V + i\dr_{\ol c}\dr_c)\wt\cH_\G
\ee
with respect to the composite graded connection $(VV\wt\G)_S$
(\ref{cmp122}) on the fiber bundle $V\Pi\to X$. Let $dV$ be a volume element
on $X$ and $\wt\cH_\G\om=FdV$, where $F$ is a function on $\Pi$. Then 
\be
\wt\cH_{S\G}\om = F_SdV=-i(\ol F_1\circ F_1 +F_1\circ \ol F_1)dV, 
\ee
where $F_1$, $\ol F_1$ and $F_S$ are the BRST and anti-BRST charges
(\ref{cmp135}) and (\ref{cmp133}). The similar splitting of a
super-Hamiltonian is the corner stone  of supersymmetric mechanics.$^{30,31}$

\end{document}